\documentclass[modern]{aastex631}

\newcommand{\tractor}{\texttt{Tractor}}
\newcommand{\thetractor}{\texttt{The Tractor}}
\newcommand{\atlas}{Siena Galaxy Atlas}
\newcommand{\leda}{HyperLeda}
\newcommand{\shortatlas}{SGA-2020}
\newcommand{\unwise}{unWISE}
\newcommand{\shortdlis}{Legacy Surveys}
\newcommand{\dlis}{DESI Legacy Imaging Surveys}

\newcommand{\diamleda}{\ensuremath{D_{\mathrm{L}}(25)}}

\submitjournal{ApJS}

\shorttitle{SGA-2020}
\shortauthors{Moustakas et~al.}

\begin{document}

\title{Siena Galaxy Atlas 2020}

\correspondingauthor{John Moustakas}
\email{jmoustakas@siena.edu}

\author[0000-0002-2733-4559]{John Moustakas}
\affil{Department of Physics and Astronomy, Siena College, 515 Loudon Road,
  Loudonville, NY 12110, USA}  

\author[0000-0002-1172-0754]{Dustin Lang}
\affil{Perimeter Institute for Theoretical Physics, 31 Caroline Street North,
  Waterloo, ON  N2L 2Y5, Canada} 

\author[0000-0002-4928-4003]{Arjun Dey}
\affil{NSF's NOIRLab, 950 N. Cherry Avenue, Tucson, AZ 85719, USA
} 

\author[0000-0002-0000-2394]{St\'ephanie Juneau}
\affil{NSF's NOIRLab, 950 N. Cherry Avenue, Tucson, AZ 85719, USA
} 

\author{Aaron Meisner}
\affil{NSF's NOIRLab, 950 N. Cherry Avenue, Tucson, AZ 85719, USA
} 

\author{Adam D.\ Myers}
\affil{Department of Physics and Astronomy, University of Wyoming, Laramie, WY
  82071, USA} 

\author[0000-0002-3569-7421]{Edward F. Schlafly}
\affil{Space Telescope Science Institute, 3700 San Martin Drive, Baltimore, MD
  21218, USA} 

\author[0000-0002-5042-5088]{David J. Schlegel}
\affil{Lawrence Berkeley National Laboratory, 1 Cyclotron Road, Berkeley, CA
  94720, USA} 

\author[0000-0001-5567-1301]{Francisco Valdes}
\affil{NSF's NOIRLab, 950 N. Cherry Avenue, Tucson, AZ 85719, USA
} 

\author{Benjamin A. Weaver}
\affil{NSF's NOIRLab, 950 N. Cherry Avenue, Tucson, AZ 85719, USA
} 

\author[0000-0001-5381-4372]{Rongpu Zhou}
\affil{Lawrence Berkeley National Laboratory, 1 Cyclotron Road, Berkeley, CA
  94720, USA} 

\begin{abstract}
We present the 2020 version of the Siena Galaxy Atlas (SGA-2020), a multi-wavelength
optical and infrared imaging atlas of 383,620 nearby galaxies. The SGA-2020 uses optical $grz$ imaging over $\approx20,000$~deg$^{2}$ from the DESI Legacy Imaging Surveys Data
Release 9 and infrared imaging in four bands (spanning $3.4$--22$~\micron$)
from the six-year unWISE coadds; it is more than 95\% complete for galaxies
larger than $R(26)\approx25\arcsec$ and $r<18$ measured at the
26~mag~arcsec$^{-2}$ isophote in the $r$-band. The atlas delivers precise
coordinates, multi-wavelength mosaics, azimuthally averaged optical surface
brightness profiles, model images and photometry, and additional ancillary
metadata for the full sample. Coupled with existing and forthcoming optical
spectroscopy from the Dark Energy Spectroscopic Instrument (DESI), the
SGA-2020 will facilitate new detailed studies of the star formation and
mass assembly histories of nearby galaxies; enable precise measurements of the local velocity field via the Tully-Fisher and Fundamental Plane relations; serve as a reference sample of
lasting legacy value for time-domain and multi-messenger astronomical events;
and more.
\end{abstract}

\section{Introduction}\label{sec:intro}

\subsection{Scientific Context}\label{sec:evolution}

Although a broad theoretical framework exists for understanding the physics of
galaxy formation, many key questions remain unanswered. In this framework, the
star formation and stellar mass assembly histories of galaxies are intimately
linked to the hierarchical buildup of dark matter halos, modulated by a
time-varying interplay between internal (secular) and external (environmental)
processes \citep[e.g.,][]{kennicutt98a, kormendy04a, blanton09a, somerville15a,
naab17a, wechsler18a}. Identifying these physical processes and their relative
importance---as a function of cosmic time---remains one of the foremost
outstanding problems in observational cosmology.

Some of the most pressing outstanding questions include:
\begin{itemize}
\item How and why does star formation in galaxies cease? Why do low-mass
  galaxies exhibit more extended star formation histories than massive
  galaxies?
\item What is the nature of inside-out galaxy formation---did galaxies
  start growing earlier in their inner parts, did they end star-formation
  earlier in those parts, or both?
\item How does feedback from active galactic nuclei, supernovae, stellar
  winds, and other effects regulate star formation?
\item What was the relative impact of mergers versus secular processes
  on the structure of present-day galaxies, including their bulges, bars,
  spirals, rings, warps, shells, and pseudobulges?
\item How are all these detailed processes affected by the relationship between
  galaxies and their host dark matter halo---for example, whether they are
  central or satellite galaxies?
\end{itemize}

\begin{figure*}[!t]
\centering\includegraphics[width=\textwidth]{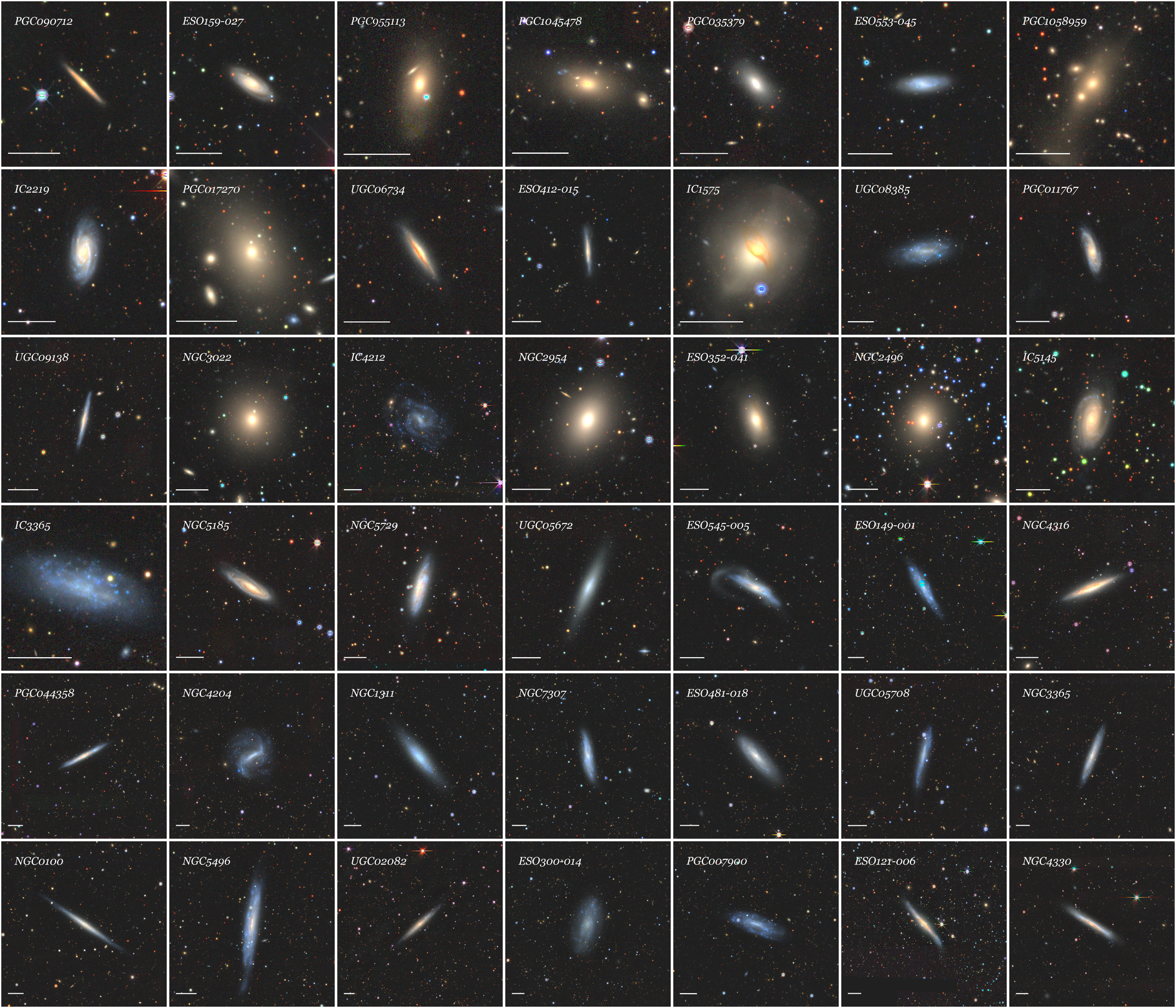}
\caption{Optical mosaics of 42 galaxies from the \shortatlas{} sorted by
  increasing angular diameter from the top-left to the bottom-right. Galaxies
  are chosen randomly from a uniform (flat) probability distribution in angular
  diameter. The horizontal white bar in the lower-left corner of each panel
  represents $1\arcmin$ and the mosaic cutouts range from $3\farcm2$ to
  $13\farcm4$. This figure illustrates the tremendous range of galaxy types,
  sizes, colors and surface brightness profiles, internal structure, and
  environments of the galaxies in the SGA. \label{fig:sample-montage}}
\end{figure*}

Investigating these and related questions requires an accurate and detailed view
of the present-day galaxy population---the population we can potentially
understand the best. In particular, galaxies which are near enough or
intrinsically large enough in terms of their apparent angular diameter to be
spatially well-resolved can be studied in significantly greater detail than more
distant, spatially unresolved galaxies.\footnote{We differentiate here between
dwarf galaxies in the Local Group whose individual \textit{stars} can be
resolved \citep{mateo98a}, and galaxies which are near or large enough to be
spatially resolved into \textit{components} (e.g., bulge vs disk) but which must
still be studied via their integrated light.} For example, in typical
ground-based optical imaging, galaxies with isophotal diameters larger than
$\approx10\arcsec-20\arcsec$ can be used to study the properties of their disk
and spheroidal components as separate, distinct features; to identify bars,
rings, disk asymmetries, and other dynamical structures; to discover and
characterize their low surface-brightness features such as stellar streams,
tidal tails, and outer envelopes; to unveil faint, low-mass satellites; and much
more.

Because of their unique and high-impact scientific potential, compilations or
atlases of large, nearby galaxies have a long, rich heritage in astronomy. In
1774, nearly 150 years before observations confirmed that the Milky Way Galaxy
was just one of many ``island universes", Charles Messier published his
\textit{Catalogue des N\'{e}buleuses et des Amas d'\'{E}toiles}, which includes
40 (now-famous) galaxies among a full catalog of 110 objects. Subsequently,
building on the naked-eye surveying effort of William Herschel and his sister
Caroline and son John \citep{herschel86a, herschel64a}, John Louis Emil Dreyer
spent more than two decades assembling the \textit{New General Catalogue of
  Nebulae and Clusters of Stars} (NGC) and the \textit{Index Catalogues} (IC), a
sample of approximately 15,000 Galactic and extragalactic objects whose
designations are still in wide-spread use today \citep{dreyer88a,
  dreyer12a}. The NGC, IC, and other early catalogs laid the foundation for
several important galaxy atlases published in the second half of the 20th
century covering most of the sky. These atlases used photographic imaging from
the Palomar Observatory Sky Survey (POSS) in the 1950s and 1960s
\citep{minkowski63a, reid93a} and the UK Schmidt Southern Sky Survey in the
1970s \citep{cannon79a}, and include: the Catalogue of Galaxies and of Clusters
of Galaxies \citep[CGCG;][]{zwicky68a}; the Uppsala General Catalog of Galaxies
and its Addendum \citep[UGC, UGCA;][]{nilson73a, nilson74a}; the Morphological
Catalog of Galaxies \citep[MCG;][]{vorontsov-velyaminov74a}; the ESO/Uppsala
Survey of the ESO (B) Atlas \citep[ESO;][]{lauberts82a}; the Principal Galaxies
Catalogue \citep[PGC;][]{paturel89a}; among others.  Eventually, data from these
and other catalogs were assembled into the indispensable Third Reference Catalog
of Bright Galaxies \citep[RC3;][]{de-vaucouleurs91a, corwin94a}. The RC3
contains extensive information on 23,011 nearby galaxies and is purportedly
complete for galaxies with $B_{\mathrm{t,Vega}}<15.5$, $D(25)>1$~arcmin, and
$v<15,000$~km~s$^{-1}$ ($z<5\times10^{-5}$), although it does include a number
of objects outside these limits which are of special interest.\footnote{$D(25)$
represents the diameter of the galaxy at the $25~\mathrm{mag~arcsec}^{-2}$
isophote in the optical and is a well-established and historically
important measure of the ``size" of a galaxy popularized by the RC3.}

The advent of wide-area ($>10^{3}$~deg$^{2}$), multi-wavelength imaging surveys
have produced the next-generation atlases of large angular-diameter galaxies in
the ultraviolet and infrared, including the 2MASS Large Galaxy Atlas
\citep{jarrett03a}, the NASA--Sloan Atlas \citep[NSA;][]{blanton05b,
  blanton11a}, the GALEX Ultraviolet Atlas of Nearby Galaxies
\citep{gil-de-paz07a}, the Spitzer Survey of Stellar Structure in Galaxies
\citep[S$^{4}$G;][]{sheth10a}, the $z=0$ Multiwavelength Galaxy Synthesis
\citep[$z0$MGS;][]{leroy19a}, and the WISE Extended Source Catalog of the 100
Largest Galaxies \citep{jarrett19a}. However, with the exception of the NSA, the
optical sizes, shapes, and total magnitudes for the objects in these atlases are
based on the photographic-plate measurements published in the RC3. The NSA,
meanwhile, delivers new optical measurements of nearby ($z<0.05$) galaxies using
$\approx8,000$~deg$^{2}$ of $ugriz$ imaging from the Sloan Digital Sky Survey
\citep[SDSS;][]{york00a}, but is $\gtrsim30\%$ incomplete at $r\lesssim14$
\citep{wake17a}.\footnote{\citet{fukugita07a} estimate the SDSS Main Survey
\citep{strauss02a} to be $\gtrsim50\%$ incomplete for $r\lesssim14$ galaxies due
to photometric shredding, which impacts other SDSS model galaxy catalogs
\citep[e.g.,][]{simard11a, meert15a}. The NSA mitigates this incompleteness to
some degree by utilizing redshifts from other surveys, but a non-negligible
fraction of the largest, brightest galaxies in the local universe are still
missed.}

\subsection{The Need for a New Large-Galaxy Atlas}\label{sec:need} 

With the historical context in mind, several recent developments motivate a
renewed effort to assemble a uniform dataset of large angular-diameter
galaxies. First, three new ground-based optical imaging surveys jointly called
the \dlis{} (hereafter, the \shortdlis) have delivered deep imaging in $g$, $r$,
and $z$ over $\approx20,000$~deg$^{2}$ of the extragalactic (i.e., low Galactic
extinction) sky (\citealt{dey19a}; Schlegel et al. 2023, in prep.). These data
provide exquisite photometric and astrometric precision and reach 1--2
magnitudes deeper than either SDSS or Pan-STARRS1 \citep{chambers16a}. In
addition, the development of the state-of-the-art image modeling code
\thetractor{} provides a computationally tractable means of working with
multi-band, multi-pass, variable-seeing imaging, enabling dedicated studies of
large angular-diameter galaxies (\citealt{lang16a}; Lang et al. 2023, in prep.).

Second, in May 2021 the Dark Energy Spectroscopic Instrument (DESI) Survey began
a five-year program (2021--2026) to obtain precise spectroscopic redshifts for
an unprecedented sample of more than 40~million galaxies and 10 million stars
over the $\approx14,000$~deg$^{2}$ \shortdlis{} footprint accessible from the
4-meter Mayall Telescope at Kitt Peak National Observatory
\citep{desi-collaboration16a, desi-collaboration16b, abareshi22a}.  As part of
this effort, the DESI Bright Galaxy Survey (BGS) is obtaining spectra and
redshifts for a statistically complete sample of $>10$ million galaxies brighter
than $r=20.175$ \citep{hahn23a}. A high-quality photometric catalog of the
largest angular-diameter galaxies over the DESI footprint is needed to ensure
the BGS has high completeness and does not suffer from the photometric shredding
and spectroscopic incompleteness of the SDSS at bright magnitudes
\citep{fukugita07a, wake17a}.

To address these and other needs, we present the 2020 version of the \atlas{}
(SGA), \shortatlas, an optical and infrared imaging atlas of nearly
$4\times10^{5}$ galaxies approximately limited to an angular diameter of
$25\arcsec$ at the $26$~mag~arcsec$^{-2}$ isophote for galaxies brighter than
$r\approx18$ over the $20,000$~deg$^{2}$ footprint of the \dlis{} Data Release 9
(LS/DR9; \citealt{dey19a}; Schlegel et al. 2023, in prep.; see
Figure~\ref{fig:sample-montage}). The \shortatlas{} delivers precise
coordinates, multi-wavelength mosaics, azimuthally averaged optical surface
brightness and color profiles, model images and photometry, and additional
metadata for the full sample.\footnote{\url{https://sga.legacysurvey.org}}
Notably, for many of the largest (e.g., NGC/IC/Messier) galaxies in the sky,
especially outside the SDSS imaging footprint, the \shortatlas{} delivers the
first reliable measurements of the optical positions, shapes, and sizes of large
galaxies since the RC3 was published more than 30 years ago.

By combining existing (archival) spectroscopic redshifts with forthcoming DESI
spectroscopy, the \shortatlas{} will spur a renewed effort to tackle several
outstanding problems in extragalactic astrophysics, particularly the interplay
between galaxy formation and dark matter halo assembly. It will also support
studies of time-domain and multi-messenger astronomical events, which are often
hampered by incomplete or heterogeneous catalogs of large, nearby galaxies which
are most likely to host the electromagnetic counterparts of gravitational wave
events \citep{gehrels16a, abbott20a} and other classes of transients. Moreover,
within the $\approx14,000$~deg$^{2}$ DESI footprint, the \shortatlas{} is
ensuring high photometric completeness for the BGS, and is facilitating
high-impact ancillary science through a variety of secondary targeting programs
\citep{myers23a, desi-collaboration23a}. For example, the DESI Peculiar Velocity
Survey will place precise new constraints on the growth rate of large-scale
structure by measuring the Tully--Fisher and Fundamental-Plane scaling relations
\citep{tully77a, djorgovski87a} from \shortatlas{} targets at $z<0.15$
\citep{saulder23a}. And finally, the \shortatlas{} will help engage the broader
public with visually striking color mosaics of large, well-resolved, nearby
galaxies, enabling a myriad of educational and public-outreach activities.

We organize the remainder of the paper in the following way: In
\S\ref{sec:sample} we define the \shortatlas{} parent sample and describe the
procedure we use to define the final sample of galaxies and their associated
(angular) group membership. In \S\ref{sec:analysis}, we describe our photometric
analysis, including how we construct the custom multi-wavelength mosaics, model
the two-dimensional images of each galaxy, and measure their azimuthally
averaged surface-brightness and color profiles. Of particular interest for some
readers may be \S\ref{sec:quicksummary}, where we validate our
surface-brightness profiles and summarize the principal \shortatlas{} data
products. In \S\ref{sec:completeness}, we quantify the completeness of the SGA
and review how it improves upon existing large-galaxy catalogs, and in
\S\ref{sec:applications} we highlight some of the exciting potential scientific
applications of the \shortatlas. Finally, \S\ref{sec:summary} summarizes the
main results of this paper and outlines some of the improvements we intend to
include in the next version of the \atlas{}.

Note that unless otherwise indicated, all magnitudes are on the AB magnitude
system \citep{oke70a} and have not been corrected for foreground Galactic
extinction. We report all fluxes in units of ``nanomaggies", where 1~nanomaggie
is the (linear) flux density of an object with an AB magnitude of
$22.5$.\footnote{\url{https://www.legacysurvey.org/dr9/description/\#photometry}}

\section{Parent Sample \& Group Catalog}\label{sec:sample}

\subsection{Building the Parent Sample}\label{sec:parent}

\begin{figure*}[!t]
\begin{center}
\includegraphics[width=1.0\textwidth]{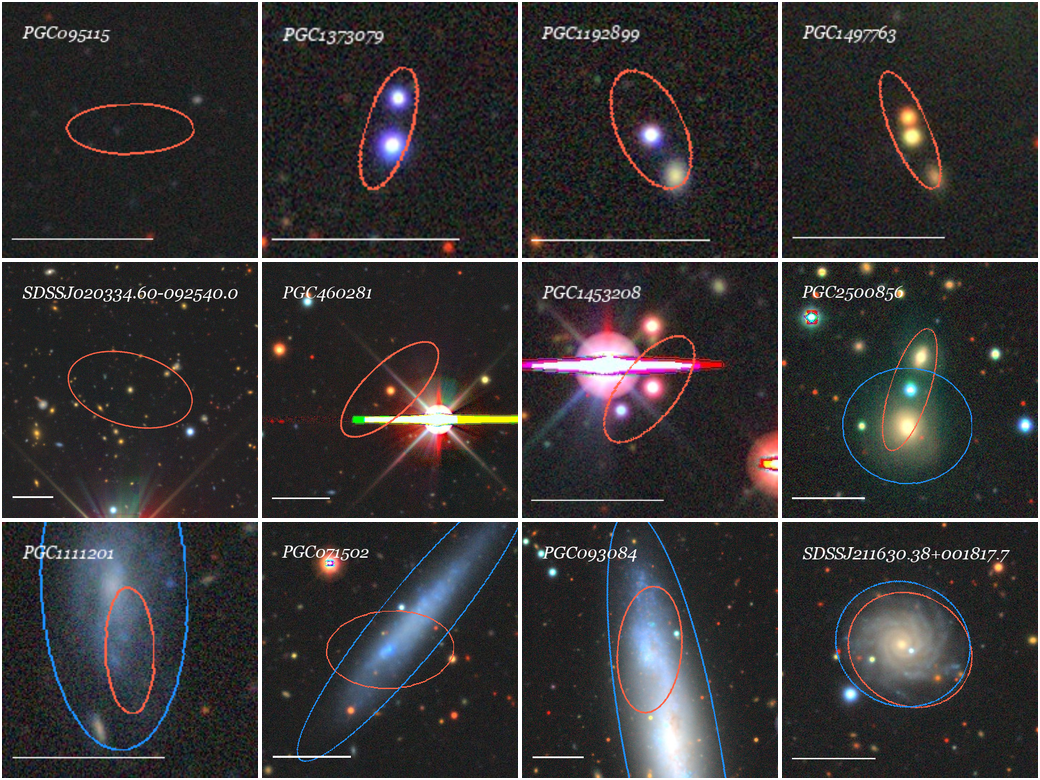}
\caption{Gallery of common types of \leda{} sources rejected while building the
  parent sample. In each panel the horizontal white bar represents
  $30\arcsec$. The top row shows one spurious object and three examples of pairs
  or triples of stars or compact galaxies which are recorded in \leda{} as one
  ``large'' galaxy (represented by the \textit{red ellipse}). The middle row
  shows four examples of how bright stars and galaxies can lead to significantly
  over-estimated galaxy diameters in \leda. For example, PGC2500856 in the
  middle-right panel is a blue star which has been miscategorized by \leda{} as
  a galaxy with a major-axis diameter of $\approx54\arcsec$ presumably due to
  the proximity of the nearby bright galaxy PGC3087062 which appears in the
  final \shortatlas{} catalog (\textit{blue ellipse}). Finally, the bottom row
  shows four examples of sources in \leda{} which are in fact photometric shreds
  or misidentified parts of a galaxy. In each case, the blue ellipse shows the
  correct parent galaxy from the \shortatlas{} while the red ellipse is the
  incorrect source from \leda{} which we remove from the parent sample (see
  \S\ref{sec:parent}). For example, PGC071502 and PGC093084 are \ion{H}{2}
  regions in ESO240-004 and NGC1507, respectively, while SDSSJ211630.38+001817.7
  is a foreground star in the body of PGC188224 with an incorrect
  angular diameter. \label{fig:rejects}}
\end{center}
\end{figure*}

Many of the largest, highest surface-brightness galaxies in the sky have been
famously known for a long time and are part of many of the legacy
(photographic-plate) large-galaxy catalogs discussed in \S\ref{sec:intro} (e.g.,
RC3). More recently, fainter, lower surface-brightness (but still ``large'',
spatially resolved) galaxies have been cataloged by modern, wide-area optical
and near-infrared imaging surveys like the SDSS, the Two Micron All Sky
Survey \citep[2MASS;][]{skrutskie06a}, and Pan-STARRS1 \citep{chambers16a}. For
the first version of the \atlas, we opt to build upon this body of previous work
by beginning from these and other catalogs of ``known'' large angular-diameter
galaxies (but see the discussion in \S\ref{sec:summary}).

Fortunately, several user-oriented databases exist which curate the positions,
sizes, magnitudes, redshifts, and other information on millions of extragalactic
sources cataloged by different surveys, including SIMBAD \citep{wenger00a}, the
NASA Extragalactic Database \citep[NED;][]{helou91a},
and \leda{} \citep{makarov14a}. After some experimentation, we opt to construct
the initial \shortatlas{} parent sample using
the \leda\footnote{\url{http://leda.univ-lyon1.fr/}} extragalactic
database. \leda{} includes extensive
metadata on nearly all known large angular-diameter galaxies, building on the
heritage of the RC3 and earlier large-galaxy atlases. In addition, an effort has
been made by the \leda{} team to \textit{homogenize} the angular diameters,
magnitudes, and other observed properties of the galaxies which have been
ingested into the database from a wide range of different surveys and
catalogs \citep{paturel97a, makarov14a}. This procedure imposes some uniformity
on our parent sample, although we show in \S\ref{sec:quicksummary} the
significant value of computing the geometry (diameter, position angle, and
ellipticity) and photometry of galaxies consistently and using modern (deep,
wide-area) optical imaging.

With these ideas in mind, we query the \leda{} extragalactic database for
galaxies with angular diameter $\diamleda>12\arcsec$ ($0\farcm2$), where
\diamleda{} is the major-axis diameter of the galaxy at the
$25$~mag~arcsec$^{-2}$ surface brightness isophote in the optical (typically the
Johnson--Morgan $B$-band; see Appendix~\ref{appendix:leda} for additional
details). Our query results in an initial parent sample of 1,436,176 galaxies.

Using visual inspection and a variety of quantitative and qualitative tests, we
cull this initial sample by applying the following additional cuts: First, we
remove the Large and Small Magellanic Clouds and the Sagittarius Dwarf Galaxy
from the sample by imposing a maximum angular diameter of
$\diamleda<180\arcmin$. These galaxies span such a large projected angular size
on the sky (many degrees) that their inclusion is outside the scope of
the \atlas{} (but see \citealt{jarrett19a}). After removing these objects, the
largest angular-diameter galaxies which remain are NGC0224=M31 and NGC0598=M33
whose \diamleda{} diameters are $178\arcmin$ and $62\arcmin$, respectively.

\begin{figure*}[!t]
\begin{center}
\includegraphics[width=1.0\textwidth]{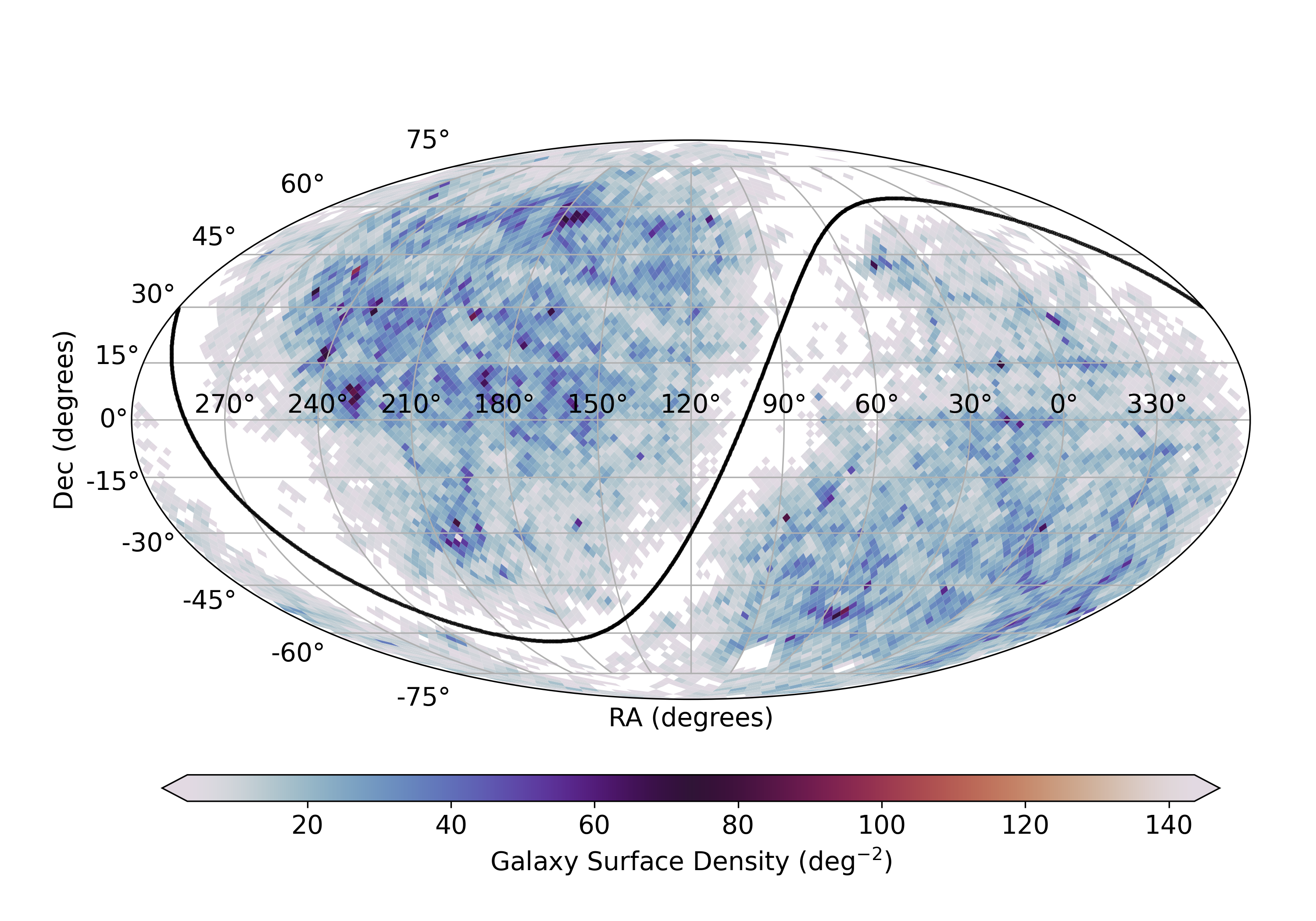} 
\caption{Distribution of 531,677 galaxies in the \shortatlas{} parent sample in
  an equal-area Mollweide projection in equatorial coordinates, binned into
  $3.4$~deg$^{2}$ healpix pixels \citep{healpy}. The dark gray curve represents
  the Galactic plane. Note the significant variations in galaxy surface density,
  which we attribute to surface-brightness incompleteness and heterogeneity in
  the aggregate \leda{} catalog. \label{fig:sgaparentsky}}
\end{center}
\end{figure*}

Furthermore, we limit the sample on the lower end to $\diamleda>20\arcsec$,
which removes roughly 900,000 galaxies (approximately 65\% of the initial
sample). We implement this cut because we find that the fraction of sources with
incorrect (usually over-estimated) diameters in \leda{} increases rapidly below
this limit. Moreover, we find that galaxies smaller than
$\diamleda\approx20\arcsec$
are well-modeled by \tractor{} as part of the standard photometric pipeline used
in LS/DR9 (\citealt{dey19a}; Schlegel et~al. 2023, in prep.). 

Next, we remove approximately 3800 galaxies with no magnitude estimate in
\leda{} (as selected by our query; see Appendix~\ref{appendix:leda}), which we
find to be largely spurious, as well as approximately 6500 objects with
significantly overestimated diameters (or spurious sources) which we identify
via visual inspection. Many of these cases are groupings of small galaxies or
stars along a line which have been misinterpreted by previous fitting algorithms
as a single edge-on galaxy. In addition, we remove approximately 1700 galaxies
whose primary galaxy identifier (in \leda) is from either SDSS or 2MASS and
whose central coordinates place it inside the elliptical aperture of another
(non-SDSS and non-2MASS) galaxy with \diamleda{} diameter greater than
$30\arcsec$. We find that in the majority of cases these objects have grossly
over-estimated diameters, presumably due to shredding by the 2MASS and SDSS
photometric pipelines. Figure~\ref{fig:rejects} displays a gallery of
some of the most common types of sources we reject from our initial parent
sample using these cuts.

\begin{figure}[!t]
\begin{center}
\includegraphics[width=1.0\textwidth]{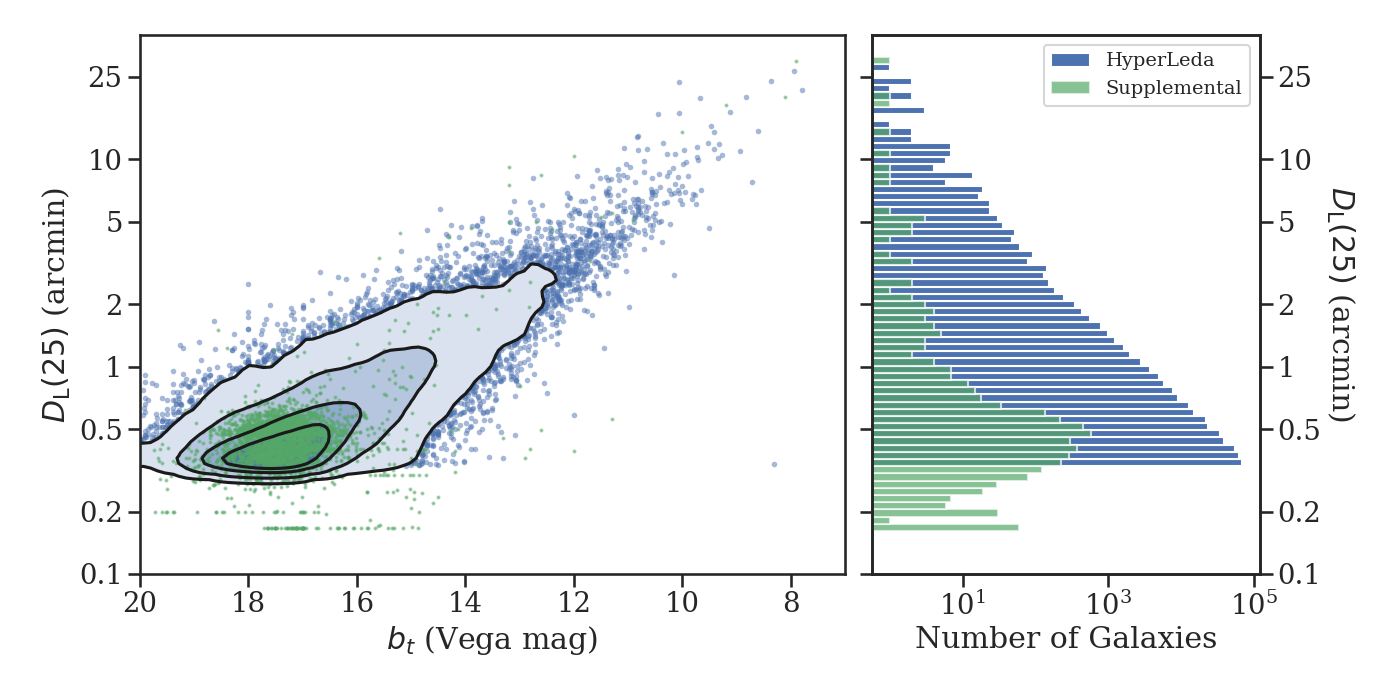}
\caption{Isophotal diameter, \diamleda, versus $b_{t}$-band magnitude
  (\textit{left}) and marginalized distribution of \diamleda{} (\textit{right})
  for the \shortatlas{} parent sample. The blue contours, points, and histogram
  represent galaxies from \leda, while the green points and histogram are
  galaxies from the supplemental catalogs we use to increase the completeness of
  the sample (see \S\ref{sec:parent}). For reference, the contours enclose 50\%,
  75\%, 95\%, and 99.5\% of the blue points. These figures show that, by
  construction, the \shortatlas{} parent sample is largely limited to
  $\diamleda>20\arcsec$ ($0\farcm333$) but with a tail of galaxies with
  diameters as small as $\approx10\arcsec$ ($\approx0\farcm167$), and it
  includes galaxies as bright as
  $b_{t,\mathrm{Vega}}\approx8$. \label{fig:d25hist}}
\end{center}
\end{figure}

Now, to improve the completeness of the parent sample, we supplement the initial
\leda{} catalog with sources drawn from three additional catalogs, making sure
to carefully handle duplicate entries. First, we add a subset of the Local Group
dwarf galaxies from \citet{mcconnachie12a}. From the original sample of 93
galaxies in \citet{mcconnachie12a}, we remove 47 which have such a low
surface-brightness and are so well-resolved that including them is beyond the
scope of the current version of the \atlas.\footnote{For reference, we remove
the following Local Group dwarfs: Andromeda I, II, III, V, VII, IX, X, XI, XII,
XIII, XIV, XV, XVI, XVII, XVIII, XIX, XX, XXI, XXII, XXIII, XXIV, XXV, XXVI,
XXVII, and XXIX; Antlia; Aquarius; Bootes I and II; Canes Venatici I and II;
Carina; Coma Berenices; Draco; Hercules; Leo IV, V, and T; Pisces II;
Sagittarius dSph; Segue I and II; Sextans I; Ursa Major I and II; Ursa Minor;
and Willman 1.} For reference, the median surface-brightness of these 47 systems
is $\mu_V=28.8$~mag~arcsec$^{-2}$, well below the surface-brightness
completeness limit of the \shortatlas{} (see
\S\ref{sec:quicksummary}). Furthermore, we remove the Fornax and Sculptor dwarf
galaxies, which are higher surface-brightness
($\mu_V=25.1-25.5$~mag~arcsec$^{-2}$) but very well-resolved into stars and star
clusters and too challenging to include in this initial version of the SGA.

Next, we add 190 galaxies from the RC3 and
OpenNGC\footnote{\url{https://github.com/mattiaverga/OpenNGC}} catalogs which
are missing from our initial \leda{} sample. Surprisingly, many of these systems
are large and have high average surface brightness; however, we suspect that an
issue with our database query (see Appendix~\ref{appendix:leda}) may have
inadvertently excluded these sources from our initial catalog. And finally, we
use the \shortdlis{} Data Release 8 (DR8) photometric catalogs to identify 2890
additional large-diameter galaxies in the \shortdlis{}
footprint.\footnote{\url{https://www.legacysurvey.org/dr8}} Specifically, after
applying a variety of catalog-level quality cuts and extensive visual
inspection, we include in our parent sample all objects (not already in our
sample) from DR8 with half-light radii $r_{50}>14\arcsec$ based on their
\tractor{} model fits.

Our final parent sample contains 531,677 galaxies approximately limited to
$\diamleda>20\arcsec$ and spanning a wide range of magnitude and surface
brightness. In Figure~\ref{fig:sgaparentsky} we show the celestial distribution
of this sample and in Figure~\ref{fig:d25hist} we show the range of apparent
$b_{t}$-band
magnitude\footnote{\url{http://leda.univ-lyon1.fr/leda/param/bt.html}} and
angular diameter spanned by the parent sample. We discuss the completeness of
the sample in \S\ref{sec:completeness}.

\subsection{Projected Group Catalog}\label{sec:groups}

Our approach for the \atlas{} is to jointly analyze galaxies which are
relatively close to one another in terms of their projected (or angular)
separation in order to properly fit overlapping light profiles. We emphasize that
we do not require galaxies to be \textit{physically} associated, which would
require knowledge of their redshifts or physical separation. We build a simple
group catalog from the parent sample described in \S\ref{sec:parent} using the
\textit{spheregroup} friends-of-friends
algorithm.\footnote{\url{https://pydl.readthedocs.io/en/latest}} We use a $10\arcmin$
linking length, taking care to ensure that galaxies assigned to the same group
overlap within two times their circularized \diamleda{} diameter.

Using this procedure, we identify 14,930 projected galaxy groups with two
members, 1585 groups with 3--5 members, 51 with 6--10 members, and just four
groups with more than 10 members, including the center of the Coma Cluster, the Virgo Cluster, and Abell~3558 (although Abell~3558 is outside the LS/DR9 imaging
footprint; see \S\ref{sec:quicksummary}). Notably, 496,255 objects or 93\% of
the parent sample, are isolated according to the criteria used to build the
group catalog.

For each galaxy group, we compute several quantities which we refer to in
subsequent sections of the paper (but see Appendix~\ref{appendix:data} for the
complete data model). \textsc{group\_name} is a unique group name, which we
construct from the name of the group's largest member (ranked by \diamleda) and
the suffix \textsc{\_group} (e.g., NGC4406\_GROUP). For isolated
galaxies, \textsc{group\_name} is just the name of its only member (i.e.,
without the \textsc{\_group} suffix). In addition, we compute \textsc{group\_ra}
and \textsc{group\_dec} to be the \diamleda-weighted right ascension and
declination, respectively, of all the group members. Once again, for isolated
systems, \textsc{group\_ra} and \textsc{group\_dec} are identical to
the \textsc{ra} and \textsc{dec} coordinates of that galaxy (see
Appendix~\ref{appendix:data}). Finally, we record our estimate of the diameter
of the group in the quantity \textsc{group\_diameter}. For isolated
galaxies, \textsc{group\_diameter} equals \diamleda, but for groups we
compute \textsc{group\_diameter} to be the maximum separation of all the pairs
of group members plus their \diamleda{} diameter (in arcmin).

\section{Photometric Analysis}\label{sec:analysis} 

\subsection{Imaging Data}\label{sec:imaging}

We build the \shortatlas{} from the same optical and infrared imaging data used to produce the \shortdlis{} DR9 (\citealt{dey19a}; Schlegel et~al. 2023, in
prep.).\footnote{\url{https://www.legacysurvey.org/dr9}} Briefly, the
optical data consist of $grz$ imaging over $\approx20,000$~deg$^{2}$ from a
number of different surveys. In the North Galactic Cap (NGC), we use data from
the Beijing–Arizona Sky Survey \citep[BASS;][]{zou17a}, which provides
$\approx5,000$~deg$^{2}$ of $gr$ imaging using the 90Prime Camera
\citep{williams04a} on the Steward Observatory Bok 2.3-meter telescope at Kitt
Peak National Observatory (KPNO); and data from the Mayall $z$-band Legacy
Survey (MzLS), which provides $z$-band imaging over the same
$\approx5,000$~deg$^{2}$ footprint as BASS using the Mosaic-3 camera
\citep{dey16a} at the KPNO Mayall 4-meter telescope.

In the South Galactic Cap (SGC) and in the NGC up to a declination of
approximately $+32\arcdeg$, we use $grz$ imaging over $\approx15,000$~deg$^{2}$
from roughly 50 distinct (but uniformly processed) datasets obtained with the
Dark Energy Camera \citep[DECam;][]{flaugher15a} at the Cerro Tololo
Inter-American Observatory (CTIO) 4-meter Blanco telescope. Note that the
majority of this DECam imaging comes from the DECam Legacy Survey
\citep[DECaLS;][]{dey19a} and the Dark Energy Survey
\citep[DES;][]{dark-energy-survey-collaboration16a, abbott21a}.\footnote{See
\citet{dey19a} for a complete list of the DECam programs used.}

We supplement the optical data with all-sky infrared imaging at $3.4-22~\micron$
from the Wide-Field Infrared Survey Explorer
\citep[WISE;][]{wright10a}. Specifically, we use the 6-year WISE plus
NEOWISE-Reactivation \citep[NEOWISE-R;][]{mainzer14a} image stacks in W1
($3.4~\micron$) and W2 ($4.6~\micron$) from \citet{meisner21a}, and the W3
($12~\micron$) and W4 ($22~\micron$) unblurred image stacks from
\citet{lang14a}; collectively, we refer to these custom image stacks as the
\unwise{} coadds.\footnote{\url{http://unwise.me}} Note that in the
\shortatlas{} we produce WISE image coadds for each galaxy (or galaxy group)
in the sample but do not measure the infrared surface-brightness profiles; however, we intend
to deliver the infrared (and ultraviolet) surface-brightness profiles and integrated
photometry in a future version of the \atlas{} (see \S\ref{sec:summary}).

\subsection{Multi-Wavelength Mosaics \& Surface Brightness
Profiles}\label{sec:ellipse}  

At this point in the analysis we have multiband optical and infrared imaging
covering roughly half the sky (\S\ref{sec:imaging}) and an input parent catalog
of central coordinates and system diameters for more than half a million sources
(\S\ref{sec:groups}). The next steps are to build custom multi-wavelength
mosaics centered on each of these positions (\S\ref{sec:mosaics}); model all the
sources in the field using \thetractor{} (\S\ref{sec:tractor}); and measure
elliptical aperture photometry and azimuthally averaged surface-brightness
profiles (\S\ref{sec:sbprofiles}).

\begin{figure*}[!t]
\centering\includegraphics[width=1.0\textwidth]{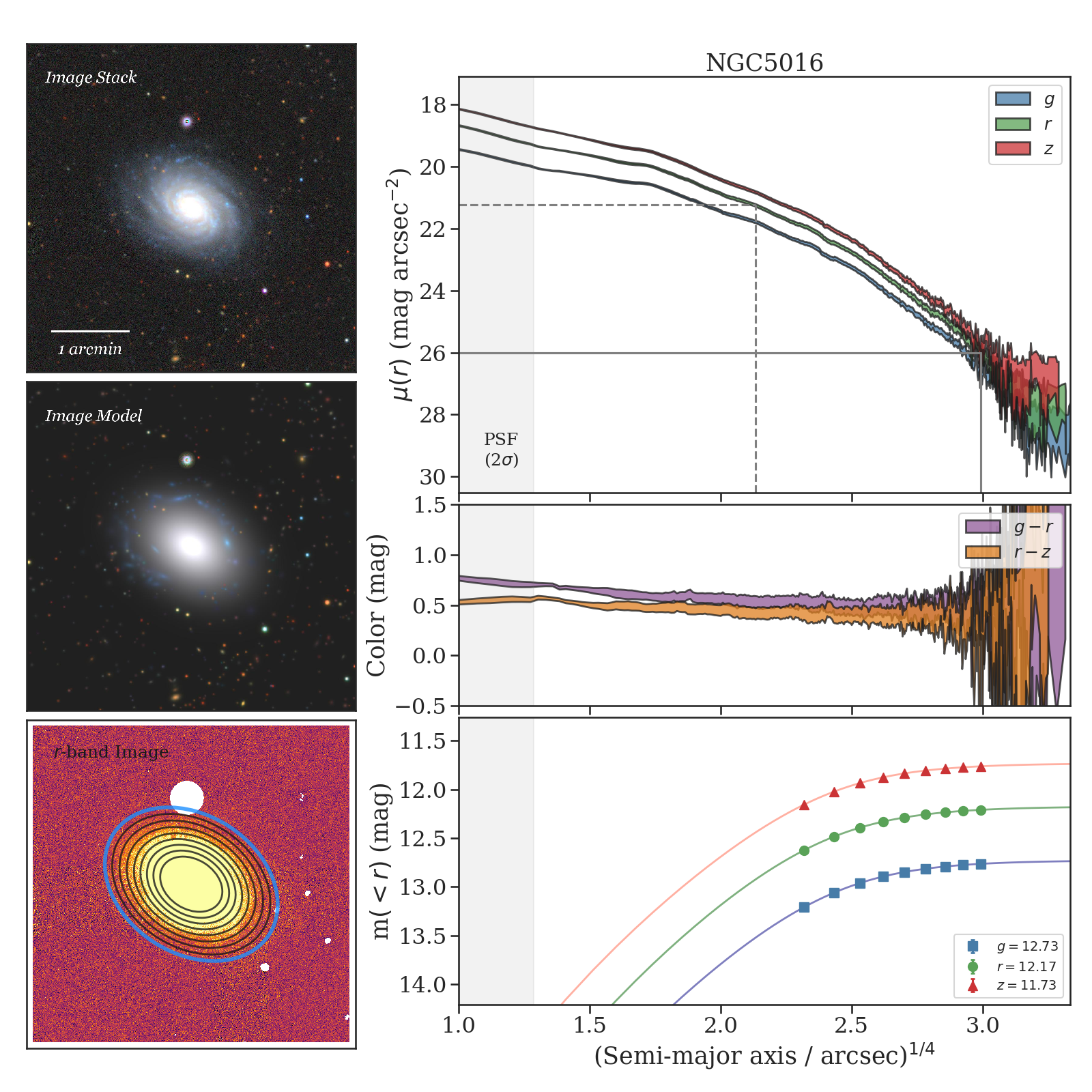} 
\caption{Illustration of the key steps and data products of the \shortatlas{}
  pipeline for one example galaxy, NGC5016. The three panels on the left-hand
  side show (\textit{top-left}) a color montage of the optical imaging;
  (\textit{middle-left}) a color montage of the corresponding \tractor{} model
  image; and (\textit{bottom-left}) the $r$-band image with masked pixels zeroed
  out (\textit{white pixel values}). The nested black ellipses in this panel
  correspond to nine surface-brightness levels between $\mu_{r}=22$ and
  $26$~mag~arcsec$^{-2}$ in $0.5$~mag~arcsec$^{-2}$ steps, with the solid blue
  isophote representing the outermost, $R(26)$, size of the galaxy. The
  right-hand panels show (\textit{top-right}) the azimuthally averaged $g$-
  (\textit{blue}), $r$- (\textit{green}), and $z$-band (\textit{red})
  surface-brightness profiles as a function of the semi-major axis;
  (\textit{middle-right}) the observed-frame $g-r$ (\textit{purple}) and $r-z$
  (\textit{orange}) color profiles; and (\textit{bottom-right}) the apparent
  brightness of NGC5016 in $g$ (\textit{filled blue squares}), $r$
  (\textit{filled green circles}), and $z$ (\textit{filled red triangles})
  measured within the same elliptical apertures shown in the lower-left
  panel. The dashed- and solid-gray lines in the top-right panel indicate, for
  reference, the $r$-band half-light radius from equation~(\ref{eq:halflight})
  and the $R(26)$ radius, respectively, and the bottom-right panel also shows
  the best-fitting curve-of-growth models (independently fit to the photometric
  data in each bandpass) given by equation~(\ref{eq:curveofgrowth}). The legend
  in the lower-right panel also provides the integrated (asymptotic) $grz$
  magnitudes for this galaxy. \label{fig:sbmontage_example1}}
\end{figure*}

Before proceeding, we briefly summarize the principal software products we use,
as well as the relationship between them. First, given an astrometrically and
photometrically calibrated image, inverse variance image, and knowledge of the
point-spread function (PSF),
\thetractor\footnote{\url{https://github.com/dstndstn/tractor}} uses the PSF and
a family of two-dimensional galaxy models to forward-model the observed
pixel-level data (\citealt{lang16a}; \citealt{dey19a}; Lang et~al. 2023, in prep.). One of the principal advantages of \thetractor{} is that it
handles multi-band, multi-CCD, variable-PSF imaging in a statistically rigorous
way, which is especially important when dealing with the full range of optical
and infrared (WISE) data from the \shortdlis. In order to handle the imaging data, the \shortdlis{} team has developed
\texttt{legacypipe}\footnote{\url{https://github.com/legacysurvey/legacypipe}},
a photometric pipeline which wraps \thetractor{} as its fitting engine and
conveniently handles many tasks related to this imaging dataset
\citep{dey19a}. Finally, for the \atlas{} project specifically, we have
developed the \texttt{SGA}\footnote{\url{https://github.com/moustakas/SGA}} and
\texttt{legacyhalos}\footnote{\url{https://github.com/moustakas/legacyhalos}}
software products, which rely on some of the lower-level \texttt{legacypipe}
functionality (with MPI-parallelization) but also include code to carry out the
non-parametric photometric analysis which is one of the cornerstone data
products of the \shortatlas{} (see \S\ref{sec:quicksummary}).

\subsubsection{Mosaics}\label{sec:mosaics}

Given the diameter of each system and its central coordinates
(\textsc{group\_ra}, \textsc{group\_dec}, and \textsc{group\_diameter}; see
\S\ref{sec:groups}), we first determine if LS/DR9 imaging exists in all three
$grz$ bands over at least 90\% of the area. If not, we remove that system from
further analysis (including, unfortunately, NGC0224=M31, where we only have MzLS
$z$-band imaging). Then, for the remaining objects, we generate $g$-, $r$-, and
$z$-band mosaics with a size which depends on the group size: for groups with
$\textsc{group\_diameter}<14$\arcmin{} we generate a mosaic of diameter
$3\times\textsc{group\_diameter}$; for groups with $14\arcmin <
\textsc{group\_diameter} < 30\arcmin$ we generate a mosaic of diameter
$2\times\textsc{group\_diameter}$; and for NGC0598=M33, whose
\textsc{group\_diameter} is $>30\arcmin$, we use a mosaic diameter of
$1.4\times\textsc{group\_diameter}$. We also choose the input imaging according
to the following criteria: We use the DECam imaging (from DECaLS and DES) for
all of the South Galactic Cap, and for the North Galactic Cap when
$\textsc{group\_dec}< 32\fdg375$; and the BASS plus MzLS imaging otherwise
\citep[see \S 4.1.3 of][]{myers23a}.

Our analysis begins with reduced and calibrated CCD-level 90Prime, Mosaic-3, and
DECam imaging. These reduced data are generated using the NOIRLab Community
Pipeline (CP) dedicated to each instrument \citep[e.g.,][]{valdes14a}, together
with several custom data-reduction steps developed by the \shortdlis{} team.  We
refer the reader to \citet{dey19a} and Schlegel et~al. (2023, in prep.)
for a detailed description of these data-reduction procedures.  Briefly, we use
Pan-STARRS1 PSF photometry \citep{chambers16a, finkbeiner16a} transformed to the
natural filter system of each instrument for photometric calibration\footnote{\url{https://www.legacysurvey.org/dr9/description/\#photometry}} and Gaia
Data Release 2 \citep{gaia-collaboration18a} stellar positions for
astrometry\footnote{\url{https://www.legacysurvey.org/dr9/description/\#astrometry}}.
The average photometric precision for bright (but unsaturated) stars is better
than $\pm10$~mmag in $grz$ over the full footprint, and the astrometric
precision is approximately $\pm0\farcs030$ for DECam and Mosaic-3 and
$\pm0\farcs12$~mas for 90Prime \citep{dey19a}.

Accurate large-galaxy photometry depends crucially on robust and unbiased
background-subtraction. For the \shortatlas, we utilize the same
background-subtracted images used for LS/DR9.  As we discuss below and in
\S\ref{sec:quicksummary}, however, the sky-subtracted data do contain systematic
errors which we intend to mitigate in future versions of the SGA (see the
discussion in \S\ref{sec:summary}).

First, the CP carefully masks astrophysical sources and then subtracts the
large-scale sky-pattern across the field of view from each exposure using a
low-order spline model derived from robust statistics measured on the individual
CCDs.  We note that the 90Prime, Mosaic-3, and DECam CCDs are approximately
$30\arcmin\times30\arcmin$, $17\farcm5\times17\farcm5$, and
$9\arcmin\times18\arcmin$, respectively, so the angular scale of this background
model is much larger than all but the largest galaxies in the \shortatlas. Next,
the CP subtracts the camera reflection pattern (or pupil ghost) from the DECam
and Mosaic-3 data and the fringe
pattern from the Mosaic-3 $z$-band and 90Prime $r$-band data. Telescope reflections from very bright stars are not removed. Next, the CP aggressively subtracts the high-frequency pattern noise (caused
by a drifting amplifier bias level) present in the Mosaic-3 imaging. The pattern
fitting was designed to preserve counts in smaller galaxies and stars but,
unfortunately, has a significant effect on the low surface-brightness, outer
envelopes of the galaxies in our sample. This pattern-noise subtraction affects
all the MzLS imaging and causes galaxies to appear \textit{too green} in the
$grz$ color mosaics (see Appendix~\ref{appendix:issues}). For DECam, we also
subtract a residual $g$-, $r$-, and $z$-band sky pattern and a $z$-band fringe
pattern from the data using median-scaled templates derived from multiple
exposures (in a given bandpass) within one or more
nights.\footnote{\url{https://www.legacysurvey.org/dr9/sky}} Finally, we remove
the spatially varying sky-background on the smallest scales by dividing each CCD
into 512-pixel boxes, computing the robust median, and using spline
interpolation to build the final background map. During this step, we mask
pixels which lie within the elliptical aperture of any galaxy in the
\shortatlas{} parent catalog (from \S\ref{sec:parent}), as well as Gaia stars
and other sources detected in each image.

Finally, with all the reduced data in-hand, we build the full-field mosaic for
each galaxy (or galaxy group) as the inverse-variance weighted sum of all the
available imaging (in each bandpass) projected onto a tangent plane using
Lanczos-3 (sinc) resampling. For the $grz$ imaging we adopt a constant pixel
scale of $0\farcs262$~pixel$^{-1}$ and for the \unwise{} mosaics we use
$2\farcs75$~pixel$^{-1}$.  The left panels of
Figures~\ref{fig:sbmontage_example1} and \ref{fig:sbmontage_example2} show, as
examples, the $grz$ color mosaics for the isolated galaxy NGC5016 and PGC193192,
a member of the PGC193199 Group, respectively.

\subsubsection{\tractor{} Modeling \& Masking}\label{sec:tractor}

We use \thetractor{} to model all the sources in a given mosaic, including the
large angular-diameter galaxies of interest. Note that all source detection and
model fitting with \thetractor{} takes place on these coadded images (triggered
by invoking the \texttt{--fit-on-coadds} option in \texttt{legacypipe}), unlike
for the standard DR9 processing in which all model fitting is done using the
unresampled CCD images jointly (see \citealt{dey19a} and Schlegel et~al. 2023, in prep.).

Before fitting, we multiply the optical inverse variance mosaics ($iv$) by a
factor of $\sqrt{iv/iv_{50}}$, where $iv_{50}$ is the median inverse variance of
the mosaic.  This rescaling down-weights the bright central regions of galaxies
even more than they already are from source Poisson noise and has the practical
effect of mitigating the tendency of \thetractor{} to fit the high-surface
brightness central region of each galaxy at the expense of its outer envelope.
In addition, we increase the threshold for detecting and deblending sources by
specifying \texttt{--saddle-fraction~0.2} and \texttt{--saddle-min~4.0} (the
default values are 0.1 and 2.0, respectively). The
\texttt{saddle-fraction} parameter controls the fractional peak height for
identifying new sources around existing sources and \texttt{saddle-min} is the
minimum required saddle-point depth (in units of the standard deviation of pixel
values above the noise) from existing sources down to new sources. We find
these options necessary in order to prevent excessive shredding and over-fitting
of the \textit{resolved} galactic structure in individual galaxies (e.g.,
\ion{H}{2} regions). Finally, \thetractor{} detects sources, creates a
segmentation map, and then uses the mean PSF of the coadd to compute the
two-dimensional, maximum-likelihood model of each source (fitting all three
$grz$ bands simultaneously) from among the following possibilities:
\texttt{PSF}, \texttt{REX}, \texttt{EXP}, \texttt{DEV}, or
\texttt{SER}.\footnote{Briefly, \texttt{REX} is a round ($\epsilon=0$)
exponential galaxy model with variable half-light radius; \texttt{EXP} and
\texttt{DEV} represent an exponential and \citet{de-vaucouleurs48a} galaxy
profile, respectively; and \texttt{SER} is a \citet{sersic68a} galaxy model (see
\citealt{dey19a} and the LS/DR9 documentation for more details).} For reference, we construct the PSF of the coadd as the inverse-variance weighted average PSF of the individual pixelized PSFs contributing to the coadd, which is sufficient given
that the galaxies we are interested in, $\diamleda\gtrsim20\arcsec$, are
significantly larger than the optical image quality,
$\mathrm{PSF}_{\mathrm{FWHM}}\approx1\arcsec-2\arcsec$.

\begin{figure*}[!t]
\centering\includegraphics[width=1.0\textwidth]{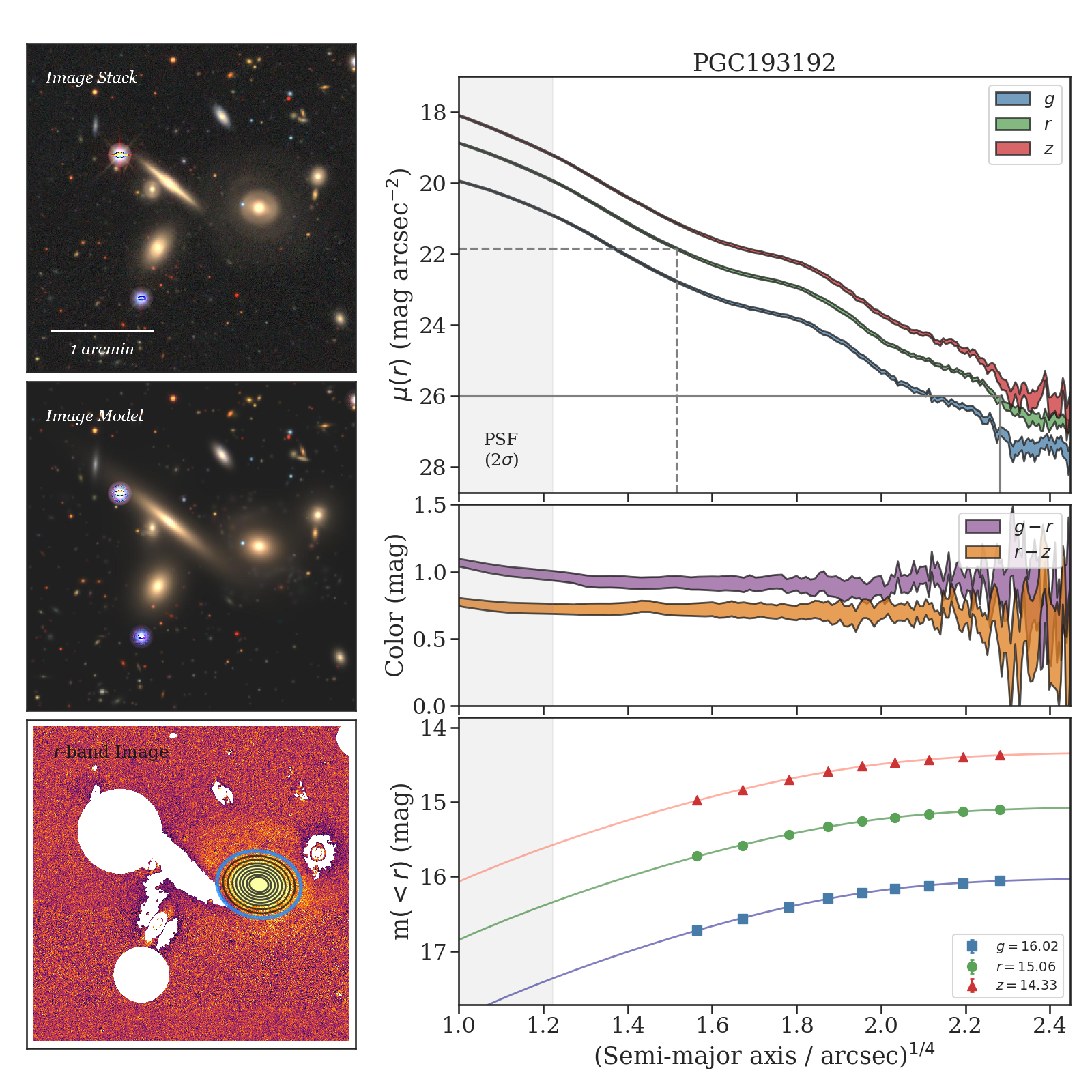}
\caption{Like Figure~\ref{fig:sbmontage_example1}, but for PGC193192, the
  second-largest member of the PGC193199 Group.
\label{fig:sbmontage_example2}}
\end{figure*}

The middle-left panels of Figures~\ref{fig:sbmontage_example1} and
\ref{fig:sbmontage_example2} show \thetractor{} model image stack for NGC5016
and the PGC193199 Group, respectively. Overall, the model is an excellent
description of the data, particularly for the small, compact sources in the
field. However, note how \thetractor{} fits the (resolved) spiral arms in
NGC5016 as elongated blue ``galaxies'', and the extended outer envelopes of the
early-type galaxy models in the PGC193199 Group compared to the data. Despite
these issues, \thetractor{} models of the large galaxies in our sample are
extremely useful and complementary to the non-parametric photometric
measurements we carry out in \S\ref{sec:sbprofiles}.

Using \thetractor{} models, we next build an image mask which we use in
\S\ref{sec:sbprofiles}. First, we read the
\texttt{maskbits}\footnote{\url{https://www.legacysurvey.org/dr9/bitmasks/\#maskbits}}
bit-mask image produced as part of the pipeline, but only retain the
\texttt{BRIGHT}, \texttt{MEDIUM}, \texttt{CLUSTER}, \texttt{ALLMASK\_G},
\texttt{ALLMASK\_R}, and \texttt{ALLMASK\_Z} bits. Hereafter, we refer to this
mask as the \texttt{starmask}. Next, we build a \texttt{residual mask} which
accounts for statistically significant differences between the data and the
\tractor{} models. In detail, we flag all pixels which deviate by more than
$5\sigma$ (in any bandpass) from the absolute value of the Gaussian-smoothed
residual image, which we construct by subtracting the model image from the data
and smoothing with a 2-pixel Gaussian kernel. This step obviously masks all
sources, including the large galaxies of interest, but we restore those pixels
in the next step. In addition, we iteratively dilate the mask two times and mask
pixels along the border of the mosaic with a border equal to 2\% the size of the
mosaic.

Then, we iterate on each galaxy in the group from brightest to faintest based on
\thetractor{} $r$-band flux and carry out the following steps: (1) For each
galaxy, we construct the model image from all \tractor{} sources in the field
\textit{except} the galaxy of interest and subtract this model image from the
data. (2) We measure the mean elliptical geometry of the galaxy (center,
ellipticity, position angle, and approximate semi-major axis length) based on
the second moment of the light distribution (hereafter, the \texttt{ellipse
  moments}) using a modified version of Michele Cappellari's
\texttt{mge.find\_galaxy}\footnote{\url{http://www-astro.physics.ox.ac.uk/~mxc/software/\#mge}}
algorithm \citep{cappellari02b}. When computing the \texttt{ellipse moments}, we
first median-filter the image with a 3-pixel boxcar to smooth out any
small-scale galactic structure and we only use pixels with
$\mu_{r}<27$~mag~arcsec$^{-2}$. (3) Finally, we combine the \texttt{residual
  mask} with the \texttt{starmask} (using Boolean logic) but we unmask pixels
belonging to the galaxy based on the \texttt{ellipse moments} geometry using 1.5
times the estimated semi-major axis of the galaxy.

Occasionally, the preceding algorithm fails in fields containing more than one
galaxy if the central coordinates of one of the galaxies is masked by a previous
(brighter) system. We consider a source to be impacted if any pixel in a
$5\times5$ pixel box centered on \thetractor{} position of the galaxy is
masked. In this case, we iteratively shrink the elliptical mask of any of the
previous galaxies until the central position of the galaxy currently being
analyzed is unmasked. We emphasize that this algorithm is not perfect,
particularly in very crowded galactic fields like the center of the Coma
Cluster, but we intend to improve it in future versions of the \atlas.  Another
occasional failure mode is if the flux-weighted position of the galaxy based on
the \texttt{ellipse moments} differs by \thetractor{} position by more than
10~pixels, which can happen in crowded fields and near bright stars and unmasked
image artifacts; in this case we revert to using \thetractor{} coordinates and
model geometry.

The bottom-left panels of Figures~\ref{fig:sbmontage_example1} and
\ref{fig:sbmontage_example2} show the final masked $r$-band image for NGC5016
and PGC193192, respectively.

\subsubsection{Surface-Brightness Profiles}\label{sec:sbprofiles}

\begin{figure*}[!t]
\begin{center}
\includegraphics[width=0.9\textwidth]{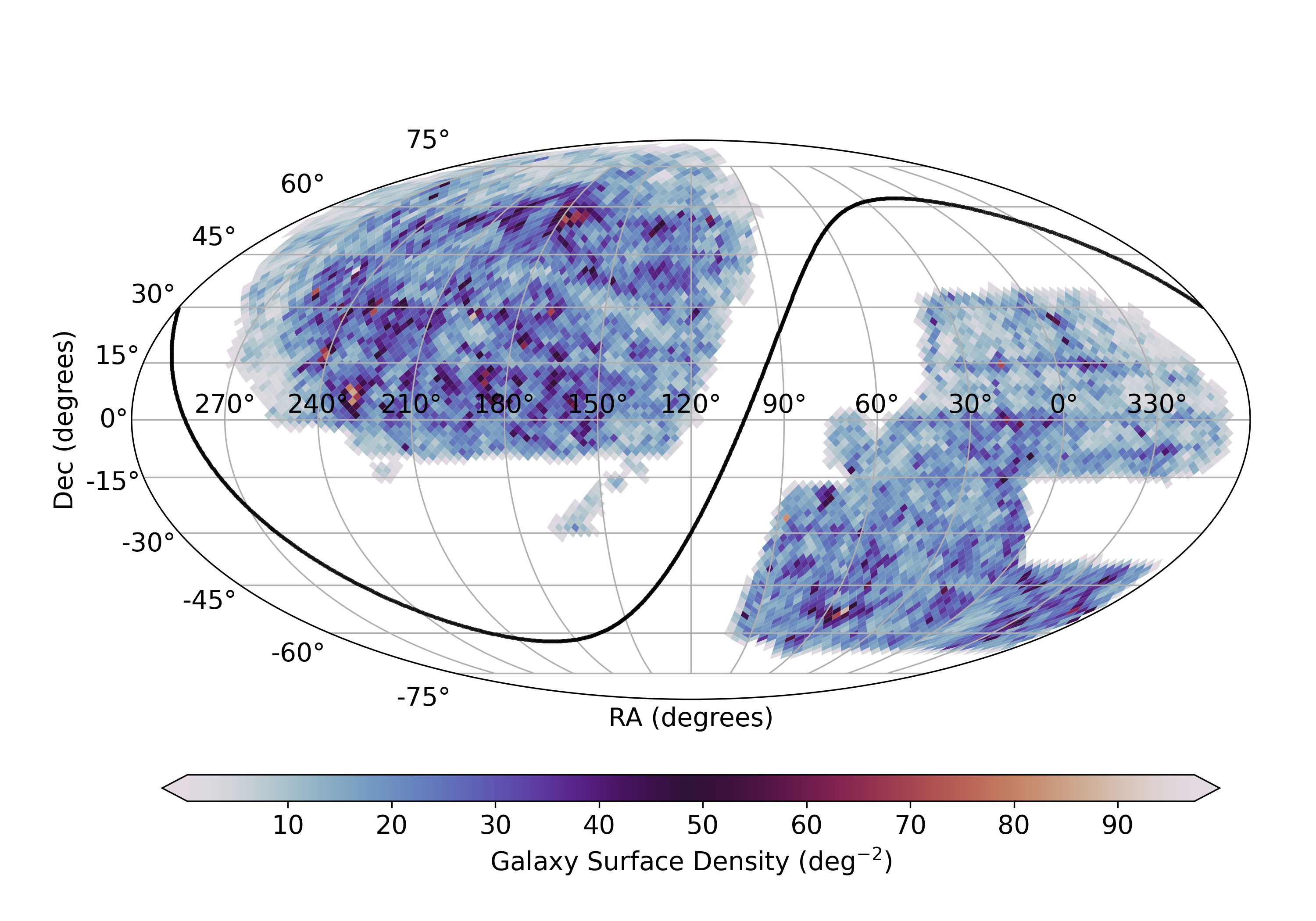}
\caption{Distribution of 383,620 galaxies in the final \shortatlas{} sample (to
  be compared with Figure~\ref{fig:sgaparentsky} but note the different colorbar
  scales). \label{fig:sgasky}}
\end{center}
\end{figure*}

With the multi-wavelength mosaics and per-galaxy image masks in-hand, we next
measure the surface brightness profiles and photometric curves of growth for
each galaxy in the sample using the standard ellipse-fitting and aperture
photometry techniques in
\texttt{photutils}\footnote{\url{https://photutils.readthedocs.io/en/stable}}
\citep{bradley23a}.  We assume a fixed elliptical geometry as a function of
semi-major axis using the \texttt{ellipse moments} measured in
\S\ref{sec:tractor}, and robustly determine the surface brightness along each
elliptical path from the light-weighted central pixel to two times the estimated
semi-major axis of the galaxy in a 1-pixel ($0\farcs262$) interval. In detail,
we measure the surface brightness (and the uncertainty) using two sigma-clipping
iterations, a $3\sigma$ clipping threshold, and median-area
integration.\footnote{In other words, we use \texttt{nclip=2}, \texttt{sclip=3},
and \texttt{integrmode=median}, as documented in the
\texttt{photutils.isophote.Ellipse.fit\_image} method.}

\begin{figure*}[!t]
\begin{center}
\includegraphics[width=1.0\textwidth]{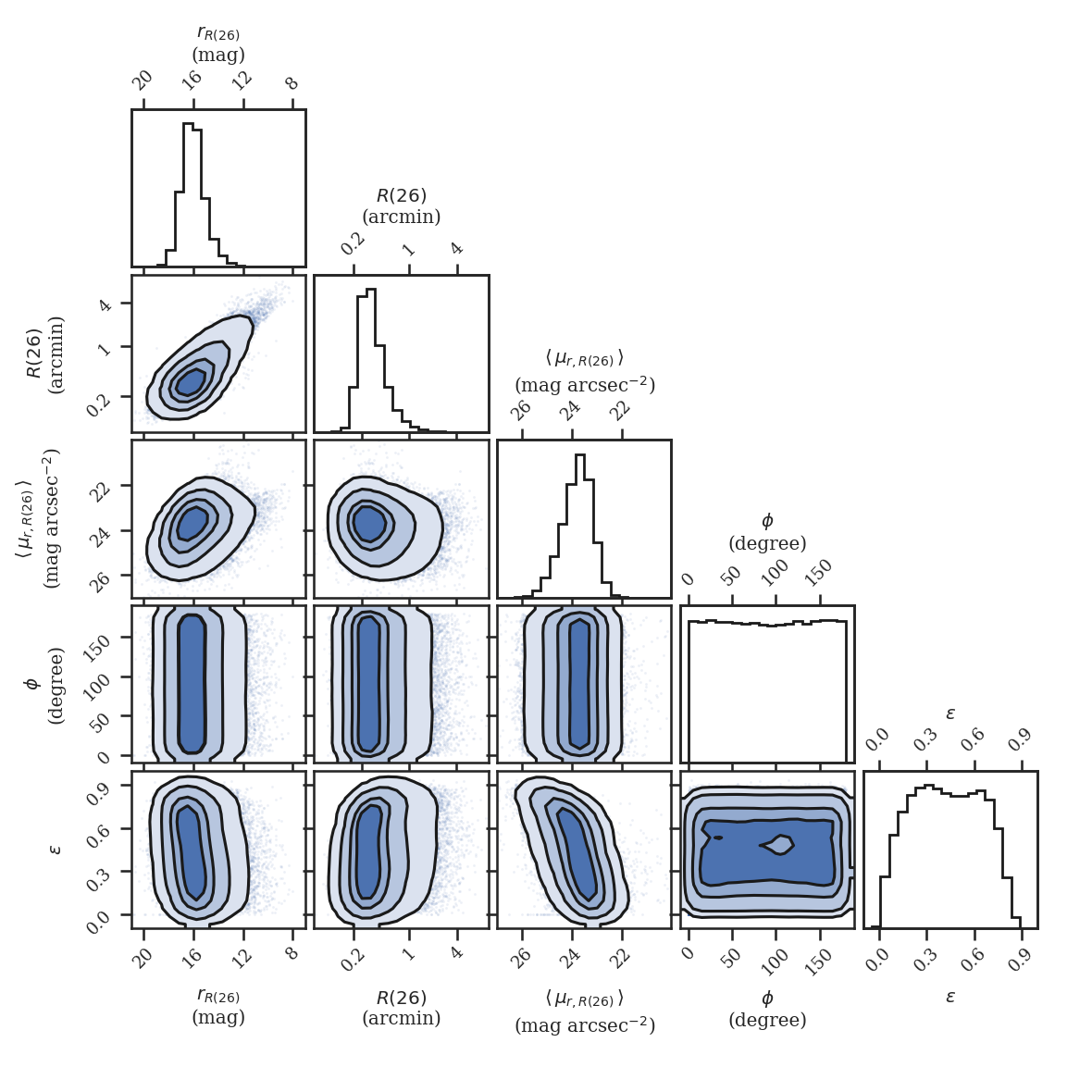}
\caption{Multivariate distribution of a subset of measured \shortatlas{} galaxy
  properties. From left to right along the bottom panels, we show $r_{R(26)}$,
  the $r$-band magnitude within $R(26)$, where $R(26)$ is the semi-major axis at
  the $26$~mag~arcsec$^{-2}$ isophote; $\langle \mu_{r,R(26)}\rangle$, the mean
  surface-brightness within $R(26)$; $\phi$, the galaxy position angle; and
  $\epsilon$, the galaxy ellipticity. The contours enclose 50\%, 75\%, 95\%, and
  99.5\% of the blue points and have been smoothed by a 0.8-pixel Gaussian
  kernel. \label{fig:properties}}
\end{center}
\end{figure*}

From the $r$-band surface brightness profile, we also robustly measure the size
of the galaxy at nine equally spaced surface-brightness thresholds between
$\mu_r=22$ and $26$~mag~arcsec$^{-2}$.  We perform these measurements by fitting
a linear model to the surface brightness profile converted to mag arcsec$^{-2}$
versus $r^{1/4}$ (which would be a straight line for a de~Vaucouleurs galaxy
profile), but only consider measurements which are within
$\pm1$~mag~arcsec$^{-2}$ of the desired surface brightness threshold. To
estimate the uncertainty in the resulting radius, we generate 30 Monte Carlo
realizations of the surface brightness profile and use the standard deviation of
the resulting distribution of radii.

We also measure the curve-of-growth in each bandpass using the tools in
\texttt{photutils.aperture}. Briefly, we integrate the image and variance image
in each bandpass using elliptical apertures from the center of the galaxy to two
times its estimated semi-major axis (based on the \texttt{ellipse moments}),
again with a 1-pixel ($0\farcs262$) interval.\footnote{Unfortunately, our
original elliptical aperture photometry had an irrecoverable bug, so in the
final release of the \shortatlas{} we infer the aperture photometry from the
surface brightness profiles; see Appendix~\ref{appendix:issues} for details.} We
fit the resulting curve-of-growth, $m(r)$, using the following empirical model:

\begin{equation}
m(r)=m_{\mathrm{tot}} + m_0 \log_{e} \left[ 1+\alpha_1
  \left(\frac{r}{r_0}\right)^{-\alpha_2} \right],
\label{eq:curveofgrowth}
\end{equation}
where $m_{\mathrm{tot}}$, $m_0$, $\alpha_1$, $\alpha_2$, and $r_0$ are constant
parameters of the model and $r$ is the semi-major axis in arcseconds. In our
analysis we take the radius scale factor $r_0=10\arcsec$ to be fixed (which
makes $\alpha_{1}$ dimensionless). Note that in the limit $r\rightarrow\infty$,
$m_{\mathrm{tot}}$ is the total, integrated magnitude. Using this model, we
infer the half-light semi-major axis length, $r_{50}$, analytically from the
best-fitting model parameters,
\begin{equation}
r_{50}= r_0 \left\{ \frac{1}{\alpha_{1}} \left[
  \exp\left(-\frac{\log_{10}(0.5)}{0.4 m_{0}}\right) - 1
  \right]\right\}^{-1/\alpha_{2}},
\label{eq:halflight}
\end{equation}

\noindent where $r_{50}$ is measured in arcseconds and $m_{0}$ is in magnitudes.

\begin{figure*}[!t]
\begin{center}
\centering\includegraphics[width=1.0\textwidth]{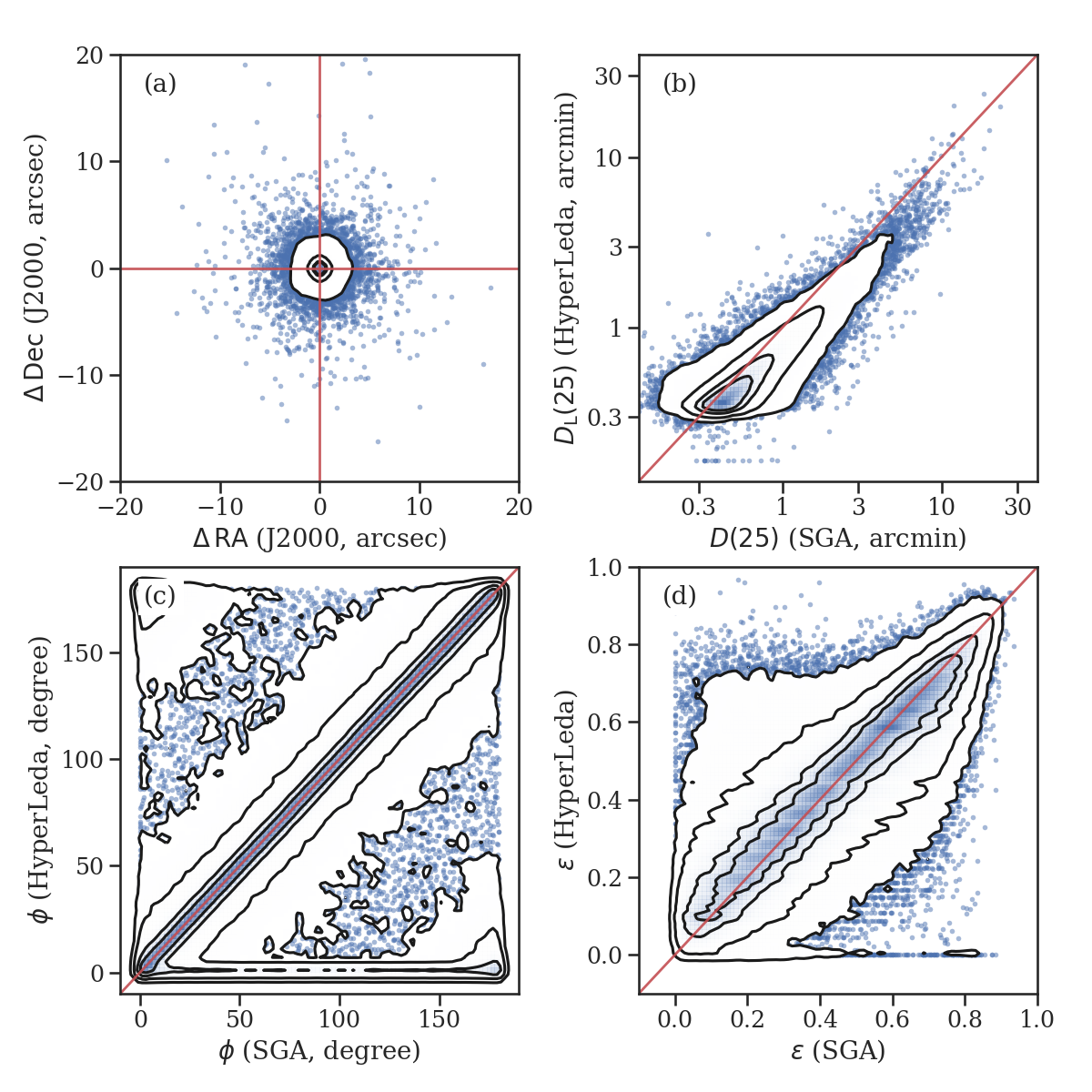}
\caption{Comparison of select observed properties reported in \leda{} against
  the newly measured quantities in the \shortatlas. (a) Difference in central
  coordinates; (b) \diamleda{} versus $D(25)$, the major-axis diameter measured
  at the $25$~mag~arcsec$^{-2}$ isophote; (c) galaxy position angle, $\phi$; and
  (d) galaxy ellipticity, $\epsilon\equiv1-b/a$, where $b/a$ is the
  minor-to-major axis ratio. In every panel, the contours enclose 50\%, 75\%,
  95\%, and 99.5\% of the blue points and have been smoothed by a 0.8-pixel
  Gaussian kernel. In panels (b), (c), and (d), the solid red line represents
  the one-to-one relation. \label{fig:sga_vs_hyperleda}}
\end{center}
\end{figure*}

\begin{figure*}[!t]
\begin{center}
\centering\includegraphics[width=1.0\textwidth]{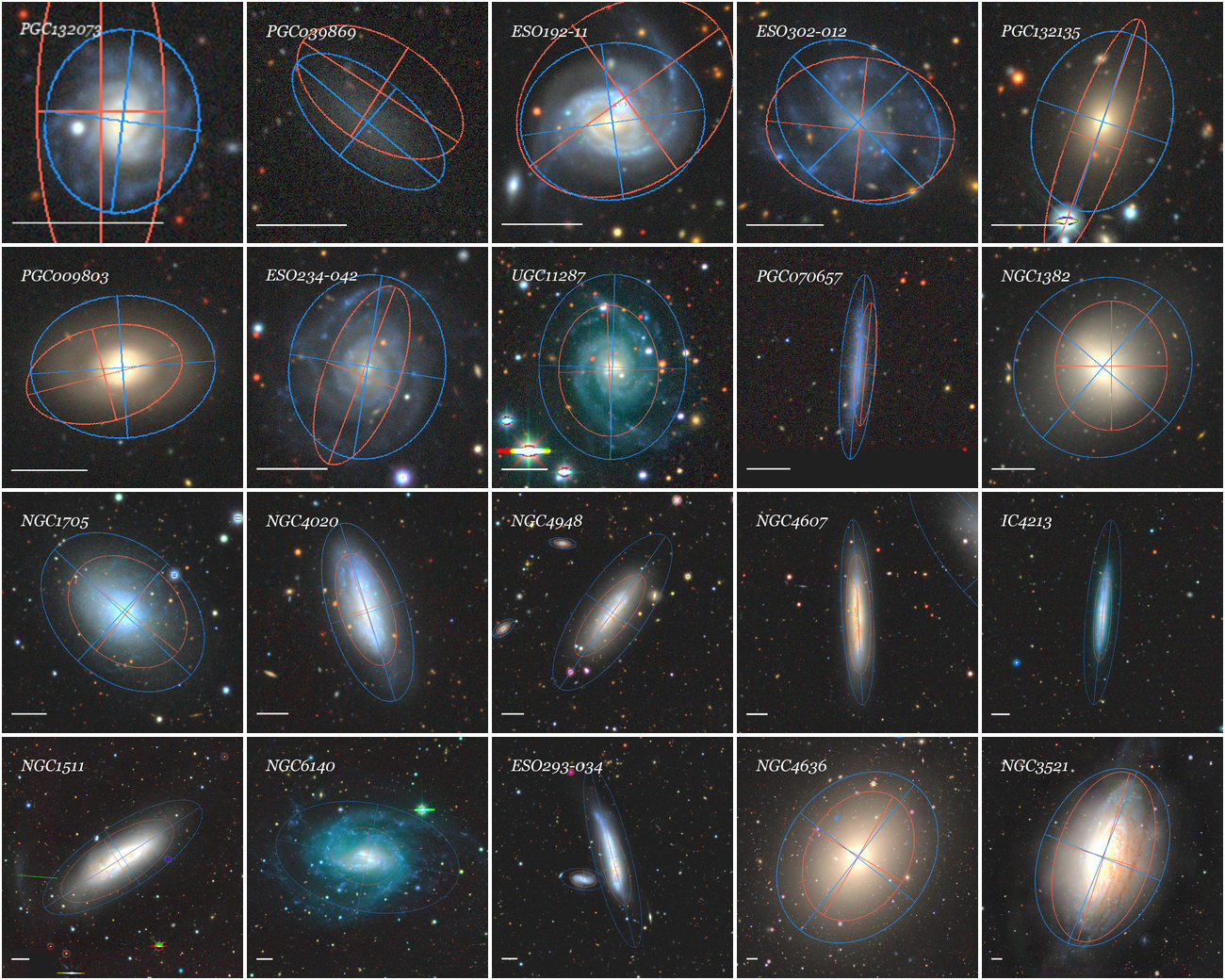}
\caption{Randomly selected gallery of 20 galaxies where the light-weighted
  central coordinates measured in the \shortatlas{} differ by more than
  $3\arcsec$ from the coordinates published in \leda, sorted by increasing
  \diamleda{} from the upper-left to the lower-right. The white bar in the
  lower-left corner of each panel represents $30\arcsec$, and the blue and red
  cross-haired ellipses represent the \shortatlas{} and \leda{} positions and
  mean geometry, respectively. Although the centers of some of these systems are
  somewhat ambiguous due to dust lanes and other irregular features (e.g.,
  NGC4607), the \shortatlas{} coordinates are generally
  superior. \label{fig:coordinates}}
\end{center}
\end{figure*}

\begin{figure*}[!t]
\begin{center}
\centering\includegraphics[width=1.0\textwidth]{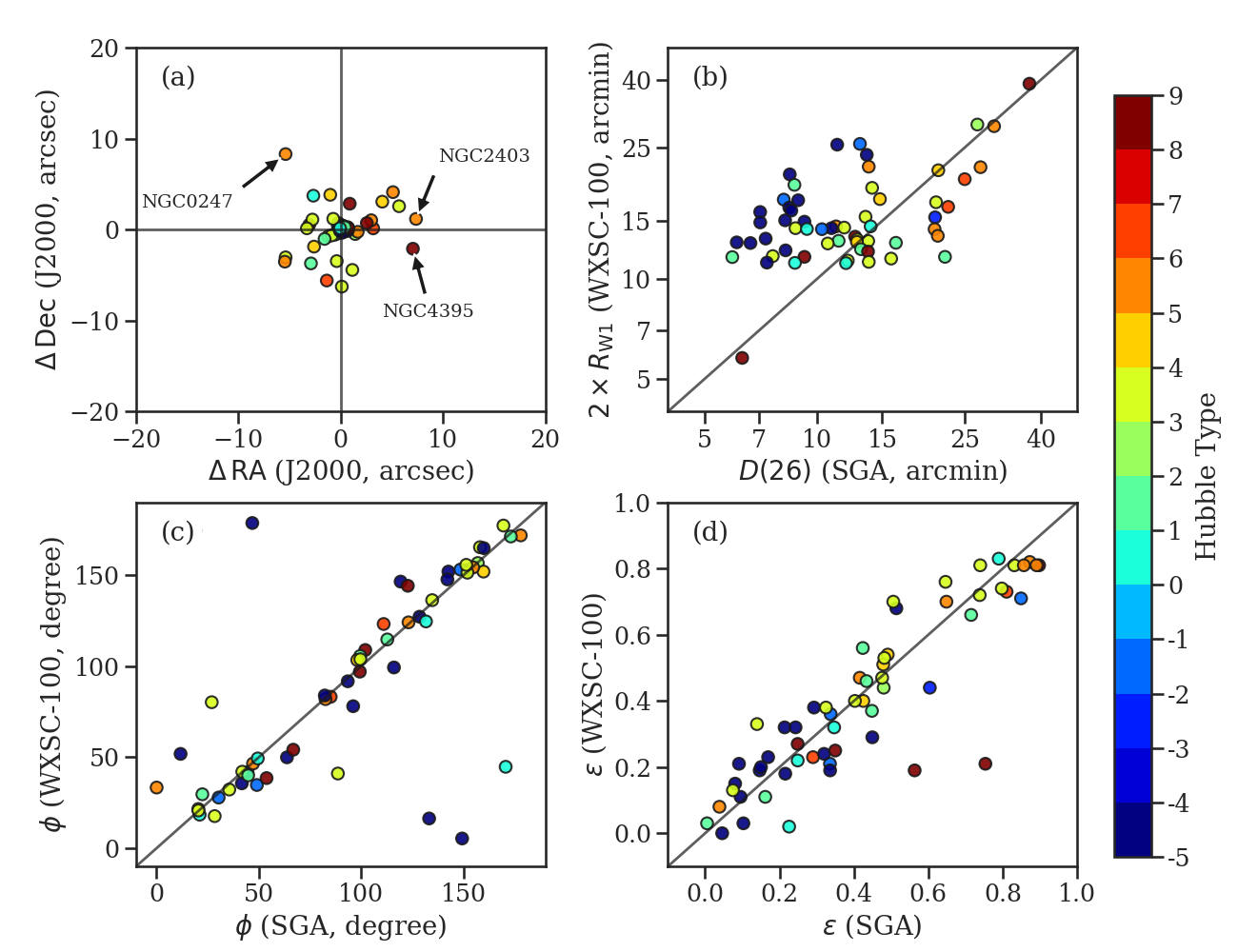}
\caption{Comparison of the (a) central coordinates; (b) diameters; (c) position
  angles, $\phi$; and (d) ellipticities, $\epsilon$ measured in the
  \shortatlas{} against those reported in the WXSC-100 \citep{jarrett19a} for an
  overlapping sample of 59 galaxies. Individual galaxies are color-coded by
  numerical Hubble type, from -5 (E) to 9 (Irr) (see
  \citealt{de-vaucouleurs91a}), as indicated by the colorbar. See the text in
  \S\ref{sec:quicksummary} for the definition of the diameter measured in the
  WXSC-100 catalog and a detailed discussion of the observed
  trends. \label{fig:sga_vs_wxsc100}}
\end{center}
\end{figure*}

\subsection{Summary \& Validation of \shortatlas{} Data Products}\label{sec:quicksummary} 

The final \shortatlas{} sample consists of 383,620 galaxies in the
$\approx20,000$~deg$^{2}$ LS/DR9 imaging footprint. The final catalog contains
precise coordinates; multi-wavelength mosaics; model images and photometry from
\thetractor; azimuthally averaged optical surface-brightness profiles; aperture
photometry and radii; and extensive metadata for all galaxies in this sample.
In this section we briefly highlight some of these measurements; for a
comprehensive description of the \shortatlas{} data products and how they can be
accessed, see Appendix~\ref{appendix:data}.

Figure~\ref{fig:sgasky} shows the celestial positions of the galaxies in the
\shortatlas{} in an equal-area Mollweide projection. The curved black line
represents the Galactic plane, which divides the sample into the North Galactic
Cap (NGC) and South Galactic Cap (SGC) imaging regions of the LS/DR9 footprint
(see \citealt{dey19a}). The total area subtended by the sample is
19,721~deg$^{2}$, covering nearly 50\% of the sky.

In Figure~\ref{fig:properties} we highlight a handful of the measured
\shortatlas{} properties, as well as the relationships between them. In a
multivariate corner plot \citep{corner}, we show $r_{R(26)}$, the $r$-band
magnitude within $R(26)$, where $R(26)$ is the semi-major axis at the
$26$~mag~arcsec$^{-2}$ isophote; $\langle \mu_{r,R(26)}\rangle\equiv r_{R(26)} +
2.5\log_{10}\,[\pi R^{2}(26)]$, the average $r$-band surface-brightness within
$R(26)$; $\phi$, the galaxy position angle (measured counter-clockwise from
North to East); and $\epsilon\equiv1-b/a$, the galaxy ellipticity, where $b/a$
is the minor-to-major axis ratio. Note the expected strong correlation between
$R(26)$ and $r_{R(26)}$, and the anti-correlation between $\epsilon$ and
$\langle \mu\rangle_{r,R(26)}$, which is due to the light in more edge-on
galaxies being attenuated more by the larger column of internal dust attenuation
(famously known as the Holmberg ``transparency test''; \citealt{holmberg58a,
  giovanelli94a}).

In Figure~\ref{fig:sga_vs_hyperleda} we compare some of the new geometrical
measurements in the \shortatlas{} against the measurements collated in \leda. In
panel (a) we plot $\Delta\mathrm{Dec}$ versus $\Delta\mathrm{RA}$, the
difference in equatorial coordinates. The overall agreement in positions is very
good; the median, mean, and $\pm1\sigma$ scatter are $0\farcs3$, $0\farcs4$, and
$\pm0\farcs5$, respectively, although the coordinates of individual (especially
irregular and low surface-brightness) galaxies differ by up to tens of
arcseconds. In Figure~\ref{fig:coordinates} we show a randomly selected set of
20 galaxies where the central coordinates in the \shortatlas{} and \leda{}
differ by more than $3\arcsec$. Although reasonable algorithms may disagree
about the central positions of some galaxies due to dust lanes or the lack of a
prominant central bulge (e.g., NGC4948=IC4156 and ESO293-034), it is clear that
the coordinates in the \shortatlas{} for most of the examples highlighted in
Figure~\ref{fig:coordinates} are more accurate in measuring the centroid of
galaxies with bright cores and the center-of-light for more diffuse systems,
even for very large, well-known galaxies like NGC4636=UGC07878 and
NGC3521=UGC06150.

Next, in Figure~\ref{fig:sga_vs_hyperleda}(b) we plot \diamleda{} (defined in
\S\ref{sec:parent}) versus $D(25)$, the \shortatlas{} major-axis diameter
measured at the $25$~mag~arcsec$^{-2}$ isophote. Not unexpectedly, we find a
strong correlation between the two quantities, although $D(25)$ is $\approx20\%$
larger, on average, than \diamleda, presumably due to the deeper optical imaging
used in the \shortatlas. Finally, in Figure~\ref{fig:sga_vs_hyperleda}(c) and
\ref{fig:sga_vs_hyperleda}(d) we plot the correlation between position angle,
$\phi$, and ellipticity, $\epsilon$, respectively, between \leda{} and the
\shortatlas. For more than 95\% of the sample, the agreement between the two
$\phi$ and $\epsilon$ measurements are excellent.

In order to validate our measurements further, in
Figure~\ref{fig:sga_vs_wxsc100} we compare the \shortatlas{} coordinates and
mean geometrical measurements against the \textit{WISE} Extended Source Catalog
of the 100 Largest Galaxies \citep[hereafter,
  WXSC-100;][]{jarrett19a}.\footnote{\url{https://vislab.idia.ac.za/research-wxsc}}
Of the 104 galaxies in the WXSC-100 sample, 59 are in the LS/DR9 imaging
footprint and consequently in the \shortatlas. Note that WISE imaging is less
affected by dust extinction and more sensitive to the underlying spatial
distribution of the older stellar population compared to our optical imaging, so
we do expect some differences in these measurements.

Focusing on panel (a) of Figure~\ref{fig:sga_vs_wxsc100} first, we find good
overall agreement in the central coordinates: the median, mean, and $\pm1\sigma$
scatter in the coordinate differences are $0\farcs76$, $2\farcs3$, and
$\pm2\farcs4$, respectively, notably smaller than the $\approx6\arcsec$~FWHM
WISE W1 point-spread function \citep{wright10a}. However, there are some notable
outliers, the three largest of which (NGC0247, NGC2403, and NGC4395) have been
annotated in the figure. Examining the optical and infrared mosaics of these and
other sources in this comparison sample, we find that the differences are due to
a combination of a lack of a distinct bright center and, for a handful of cases,
errors in the \shortatlas. For example, both NGC2403 and NGC0247 are late-type
spirals (Hubble type 6 or SABc; see \citealt{de-vaucouleurs91a}) without a
clear, bright nucleus, so the differences are arguably defensible. On the other
hand, for NGC4395 and a handful of other objects, the \shortatlas{} central
coordinates are likely incorrect (by up to $\approx7\arcsec$) relative to those
published in the WSXC-100 (see \S\ref{sec:summary} for additional discussion).

Turning next to Figure~\ref{fig:sga_vs_wxsc100}(b), we compare $D(26)$ in the
\shortatlas{} to $2\times R_{W1}$ in the WXSC-100; two times the radius of the
galaxy measured down to an isophotal level of
$\mu_{W1,AB}\approx25.7$~mag~arcsec$^{-2}$. We find the two sizes to be
reasonably well-correlated, despite the differences in imaging effective
wavelength, albeit divided into two broad sequences. The sizes of the spirals
(Hubble types 2 to 3---Sab to Sb) and later generally follow the one-to-one
relation (median difference of $-0.004$~dex) with a scatter of $0.11$~dex
($\pm30\%$), while the earlier-type, spheroidal galaxies (Hubbles types -5 to
1---E to Sa) are offset to larger sizes in the WXSC-100 by $-0.25$~dex
($\approx70\%$) with a scatter of $0.16$~dex ($\pm50\%$). As discussed by
\citet{jarrett19a}, the WISE W1-band is very sensitive to light from evolved
stars down to low surface-brightness levels (particularly due to its large
pixels, $2\farcs75$), so it is not surprising for the WSXC-100 (infrared)
diameters to be notably larger than the (optical, $r$-band) $D(26)$ diameters in
the \shortatlas. Moreover, the extended, low surface-brightness envelopes of
massive spheroidal galaxies are especially prone to over-subtraction
\citep[e.g.,][]{blanton11a, bernardi13a}, which may also be contributing to the
systematically smaller sizes of the early-type galaxies in the \shortatlas{}
relative to the WSXC-100 (see also \S\ref{sec:summary}).

Finally, in panels (c) and (d) of Figure~\ref{fig:sga_vs_wxsc100}, we compare
the position angles and ellipticities reported in the \shortatlas{} and WXSC-100
catalogs, respectively, and find very good agreement: the mean differences are
$\phi_{\mathrm{SGA}}-\phi_{\mathrm{WXSC}}=0\fdg83\pm7\fdg7$ and
$\epsilon_{\mathrm{SGA}}-\epsilon_{\mathrm{WXSC}}=0.017\pm0.12$.

\section{Catalog Completeness}\label{sec:completeness}

\begin{figure}[!t]
\begin{center}
\includegraphics[width=1\textwidth]{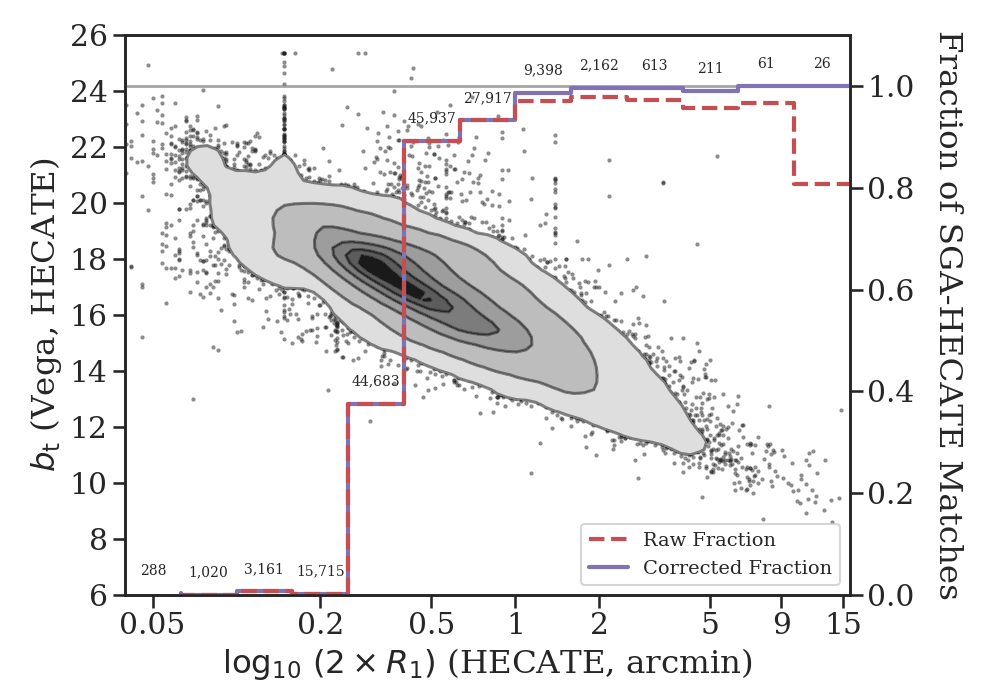}
\caption{$b_{t}$-band magnitude versus $2\times R_{1}$ for 154,093 objects in
  the HECATE value-added galaxy catalog \citep{kovlakas21a} which lie within the
  LS/DR9 imaging footprint, where $R_1$ is the semi-major axis length in
  HECATE. For reference, the contours enclose 10\%, 25\%, 50\%, 75\%, 95\%, and
  99\% of the points. The dashed-red and solid purple histograms (corresponding
  to the right-hand vertical axis) represent, respectively, the raw and
  corrected fraction of HECATE galaxies which match a source in the
  \shortatlas{} galaxy as a function of angular diameter. Finally, the small
  numbers written above each bin of the histograms report the number of HECATE
  galaxies in that $0.2$-dex wide bin. \label{fig:completeness}} 
\end{center}
\end{figure}

Quantifying the completeness of the \shortatlas{} is difficult because the
parent sample is largely defined by \leda{} (see \S\ref{sec:parent}), which
aggregates data from many different surveys---each with their own potentially
complicated selection function---from the ultraviolet to the radio. This
heterogeneity can be clearly seen in Figure~\ref{fig:sgasky} as variations in
the sample surface density on both large and small angular scales and between
the North and South Galactic Cap. Indeed, as we discuss in \S\ref{sec:summary},
one of the primary goals of the next version of the SGA is to redefine the
parent sample using the \shortdlis{} imaging itself, in order to begin with a
more uniform and quantifiable selection function over the full
area. Nevertheless, we can still assess the SGA's completeness by comparing
against existing catalogs which include large angular-diameter galaxies.

We choose to characterize the completeness of the \shortatlas{} by comparing
against HECATE \citep[version
  1.1;][]{kovlakas21a}.\footnote{\url{https://hecate.ia.forth.gr}} HECATE is an
all-sky value-added catalog of $\approx200,000$ galaxies at $z<0.047$
($\lesssim200$~Mpc) which contains a breadth of both observed and derived
(physical) galaxy properties. Although HECATE also uses \leda{} to define its
parent sample, the comparison is still valuable because the analysis carried out
by \citet{kovlakas21a} is independent of ours, and it includes additional
completeness checks against NED and against the local $B$-band luminosity
function.

First, from the parent HECATE sample of 204,733 galaxies, we identify 154,093
objects (75\%) to be within the LS/DR9 imaging footprint using the LS/DR9 random
catalogs, which contain a wealth of information about the imaging data (bandpass
coverage, depth, PSF size, etc.) at random positions over the footprint
\citep{myers23a}.\footnote{\url{https://www.legacysurvey.org/dr9/files/\#random-catalogs-randoms}}
Specifically, we merge together five random catalogs to achieve an effective
source density of 12,500~deg$^{-2}$, and conservatively retain all HECATE
objects whose center lies within $2\arcmin$ of one of these points. Next, we
remove 2844 galaxies without any size information in HECATE; visually
inspecting a random subset of these reveals that they are predominantly small
objects, $D(25)\ll30\arcsec$, well under the \shortatlas{} angular diameter
limit (Figure~\ref{fig:d25hist}). Finally, from the remaining 151,249 objects,
we match 95,800 (63\%) of them to an object in the \shortatlas{} using the
\leda{} PGC number \citep{paturel89a, makarov14a}, which both HECATE and the
\shortatlas{} record (where it is defined). We choose to use the PGC designation
because the differences in coordinates can be significant, such that a single
matching radius results in a non-negligible number of false-positives. For
example, among the PGC-matched samples, the mean difference in coordinates is
$0\farcs50\pm0\farcs92$, but with a tail which extends out to $80\arcsec$ for
IC2574, a well-known $18\farcm7$-diameter late-type (SABm) galaxy with no
well-defined center. In fact, 126 PGC-matched galaxies have coordinate
differences larger than $10\arcsec$, 90\% of which have
$D(26)>1\arcmin$. Nevertheless, with these caveats in mind, we can still match
an additional 178 objects using a $3\arcsec$ matching radius, resulting in a
final matched sample of 95,978 galaxies.

In Figure~\ref{fig:completeness} we plot $b_{t}$-band magnitude versus $2\times
R_1$ for the 154,093 HECATE galaxies which overlap the LS/DR9 footprint, where
$R_1$ is the semi-major axis length reported in HECATE. The dashed red histogram
(corresponding to the right-hand axis), shows the \textit{raw} fraction of
matching HECATE-SGA galaxies as a function of galaxy size, in uniform $0.2$-dex
wide bins of angular diameter between $\approx2\arcsec$ and
$\approx15\farcm8$. The small numbers written above each bin of the histogram
report the number of HECATE galaxies in that bin. In addition, the horizontal
gray line indicates, for reference, a matching fraction of 100\%.

Taken at face value, the (raw) matching fractions shown in
Figure~\ref{fig:completeness} are surprisingly poor. For example, these results
suggest that the \shortatlas{} is missing $3\%-5\%$ of galaxies with angular
diameters between $2\farcm5$ and $10\arcmin$ and a whopping 20\% of the
galaxies between $10\arcmin$ and $16\arcmin$. In total, we find 2.94\%
(367/12,487) of the HECATE galaxies with $2\times R_{1}>1\arcmin$ to be missing
from the \shortatlas. To explore this purported incompleteness, we visually
inspect the LS/DR9 imaging at the position of the 367 ``missing" galaxies and
find the following results: 20 objects are Local Group dwarf galaxies which
are intentionally excluded from the \shortatlas{} (see \S\ref{sec:parent}); 44
objects fall on the edge of the imaging footprint or other serendipitous (but
unfortunate) gaps in three-band ($grz$) coverage; 64 galaxies are real but the
angular diameters reported in HECATE are overestimated, sometimes by a
significant factor; 78 are part of a larger galaxy (e.g., \ion{H}{2} regions)
and other kinds of photometric shreds; and 24 are entirely spurious. The solid
purple histogram in Figure~\ref{fig:completeness} shows the corrected fraction
of HECATE-SGA matches after accounting for these errors. In the end, we find
that just 137 out of 12,257 (1.12\%) HECATE galaxies with $2\times R_{1}>1\arcmin$
are real and genuinely missing from the \shortatlas.

Although the fraction of missing galaxies is relatively low, since both the
\shortatlas{} and HECATE ultimately originate (in large part) from \leda, it is
surprising that \textit{any} objects are missing from the \shortatlas,
especially ones with angular diameters larger than one arcminute. Although we do
not know why these objects do not appear in our parent \leda{} catalog, we
suspect that an issue with the database query may be ultimately responsible (see
Appendix~\ref{appendix:leda}). In any case, we intend to ensure that these and
other missing objects serendipitously identified by the SGA team through visual
inspection are included in the next version, as we discuss in
\S\ref{sec:summary}.

What about the completeness of the \shortatlas{} among smaller angular-diameter
galaxies, $25\arcsec\lesssim 2\times R_{1} < 1\arcmin$? According to
Figure~\ref{fig:completeness}, the completeness remains relatively high, above
$\approx90\%$. However, our analysis of the $>1\arcmin$-diameter galaxies
reveals that more than half of the HECATE objects missing from the
\shortatlas{} either have incorrect (or overestimated) diameters, or are
spurious, and we have checked that these and other effects increase steeply with
decreasing angular diameter (see, for example,
Figure~\ref{fig:rejects}). Therefore, we conclude that the \shortatlas{}
completeness is likely $\gtrsim95\%$ for galaxies with angular diameters between
$\approx25\arcsec$ and $1\arcmin$.

None of this discussion, of course, addresses the surface-brightness
incompleteness of the sample, since both HECATE and the \shortatlas{} inherit
whatever incompletenesses and heterogeneities are present in \leda, which
aggregates data from many different surveys. For example, regions of the sky
which have been imaged by the SDSS \citep{strauss02a, blanton11a} or DES
\citep{dark-energy-survey-collaboration16a, abbott21a} have uniform, deep
optical imaging ($\mu_{r,50}<24.5$ and $\mu_{r}<25.6$~mag~arcsec$^{-2}$ in the
SDSS and DES, respectively, where $\mu_{r,50}$ is the $r$-band half-light
surface brightness and the DES surface brightness is measured in a $1\farcs95$
diameter aperture), but these surveys cover just $34\%$ (SDSS;
$14,000$~deg$^{2}$) and $12\%$ (DES; $5,000$~deg$^{2}$) of the sky. Other
optical and near-infrared surveys like 2MASS \citep{jarrett00a, skrutskie06a}
and Pan-STARRS1 \citep[PS1;][]{chambers16a} cover all or nearly all of the sky
(100\% and 75\% for 2MASS and PS1, respectively); however, 2MASS is relatively
shallow compared to these other surveys ($\mu_{r}\approx22.7$~mag~arcsec$^{-2}$
assuming a median $r-K_{s}\approx2.7$ color for low-redshift galaxies;
\citealt{jarrett19a}) while the photometry of bright, large angular-diameter
galaxies in PS1 is known to be problematic \citep{magnier20a, makarov22a}. In
other words, it is difficult to fully assess the incompleteness of the
\shortatlas{} given the variations in the completeness of the surveys which
contribute to \leda.

Nevertheless, we can still make some quantitative statements using the
results shown in Figure~\ref{fig:properties} and the comparisons with HECATE,
above. Based on the correlation between the mean surface-brightness, $\langle
\mu_{r,R(26)}\rangle$, and the apparent brightness within the
$26$~mag~arcsec$^{-2}$ isophote, $r_{R(26)}$, we conclude that the \shortatlas{}
is approximately 95\% complete for galaxies with $R(26)\gtrsim25\arcsec$,
$r_{R(26)}\lesssim18$, and $\langle
\mu_{r,R(26)}\rangle\lesssim26$~mag~arcsec$^{-2}$; in addition, the \shortatlas{} is
more than 99\% complete for galaxies larger than $1\arcmin$ and brighter than
$r_{R(26)}\lesssim16$ down to the same surface-brightness limit.

\section{Scientific Applications}\label{sec:applications} 

\begin{figure*}[!t]
\begin{center}
\includegraphics[width=1\textwidth]{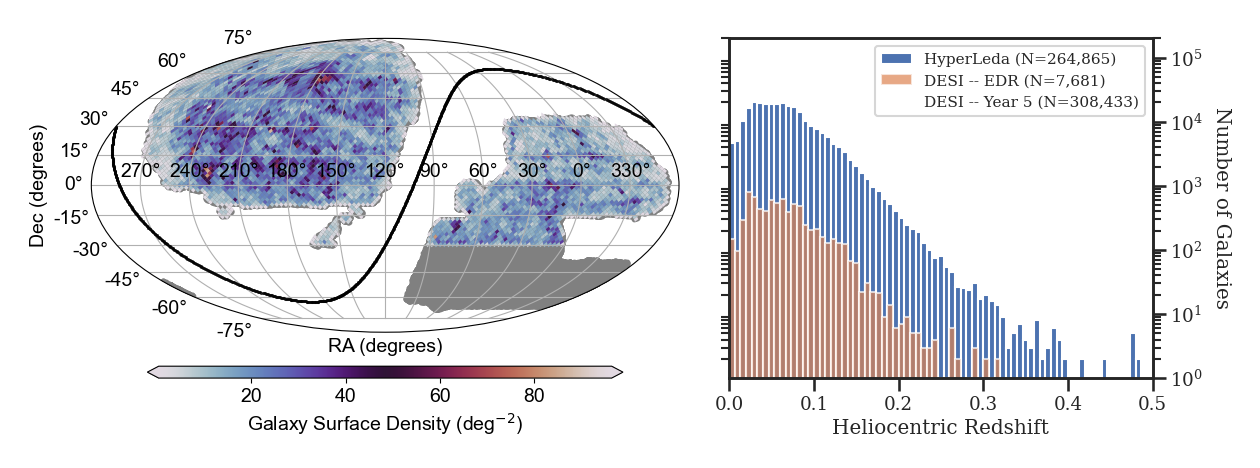}
\caption{(\textit{Left}) Celestial distribution of galaxies in the \shortatlas{}
  which are also DESI targets (\textit{colorbar}) covering
  $\approx14,000$~deg$^{2}$ \citep{myers23a} compared to the full
  $\approx20,000$~deg$^{2}$ footprint of the \shortatlas{} (\textit{dark gray
    region} in the SGC). (\textit{Right}) Redshift distribution of galaxies in
  the \shortatlas{} with existing spectroscopic redshifts from \leda{}
  (\textit{blue histogram}) and from the DESI Early Data Release (EDR;
  \textit{orange histogram}). The legend in this figure also indicates the
  approximate final number of \shortatlas{} galaxies which DESI will observe by
  the end of the five-year Main Survey. \label{fig:redshifts}}
\end{center}
\end{figure*}

As discussed in the introduction, atlases of large angular-diameter galaxies
have played a pivotal role in observational cosmology and in our modern
understanding of galaxy astrophysics and the galaxy-halo connection.  By
delivering a carefully constructed catalog of known ``large'' galaxies with new
deep optical and infrared imaging from the \dlis, we anticipate the
\shortatlas{} to play a commensurately high-impact role in a wide range of
observational studies of large, nearby, well-resolved galaxies.

The growing spectroscopic dataset from DESI is an especially powerful complement
to the \shortatlas. As discussed in \S\ref{sec:need}, DESI is targeting the
\shortatlas{} sample over $14,000$~deg$^{2}$ (70\% of the LS/DR9 footprint) as
part of the flux-limited ($r<20.175$) Bright Galaxy Survey
\citep[BGS;][]{hahn23a}.\footnote{In detail, BGS is observing all galaxies
brighter than $r=19.5$ (BGS Bright) and a color-selected subset of galaxies with
$19.5<r<20.175$ (BGS Faint); the color selection is tuned to ensure good
redshift success at this flux level in bright sky conditions.} In addition to
providing precise spectroscopic redshifts, the spectral coverage
($3600-9800$~\AA), instrumental resolution ($\mathcal{R}\approx2000-5000$), and
spectrophotometric precision ($\pm2\%$) of the DESI spectra \citep{abareshi22a,
  guy23a} will yield important insights into the physical conditions and stellar
populations of the central regions of these systems (for recent reviews, see
\citealt{conroy13a}, \citealt{kewley19a}, \citealt{sanchez20a}, and references
therein).

Figure~\ref{fig:redshifts} illustrates the tremendous scale of the DESI
dataset. Although the majority of galaxies in the \shortatlas{} have previously
measured redshifts in \leda{} ($264,865$ galaxies, or approximately 70\% of the
sample), these redshifts come from a wide range of different surveys spanning
many decades. By the end of its five-year Main Survey (2021-2026), DESI will
produce a homogeneous, high-precision spectrophotometric dataset for more than
$300,000$ \shortatlas{} targets over $14,000$~deg$^{2}$ as part of a larger
sample of approximately 14 million BGS targets and more than 25 million fainter
extragalactic targets \citep{myers23a, desi-collaboration23b}.  Indeed, in its
June 2023 Early Data Release (EDR), DESI has already delivered high-quality
spectroscopy for nearly 7700 \shortatlas{} targets which were observed during
the DESI Survey Validation period
\citep{desi-collaboration23a}.\footnote{\url{https://data.desi.lbl.gov/doc/releases/edr}}

In addition to these Main-Survey observations, one of the especially exciting
and synergistic \textit{secondary} DESI programs is the Peculiar Velocity survey
\citep{saulder23a}.\footnote{Secondary programs are bespoke scientific programs
which utilize whatever spare fibers may be available in a given DESI tile or
pointing of $\sim$5000 fibers (see \citealt{myers23a} for details).} This
program aims to use the Fundamental Plane and Tully--Fisher relations as direct
distance indicators in order to map the peculiar velocity field at $z<0.15$;
this map will be used to place new, stringent constraints on the
cosmological parameters and the growth of large-scale structure
\citep[e.g.,][]{strauss95a}. As part of this effort, galaxies in the
\shortatlas{} are being targeted not only in their bright nucleus but also at
various positions along their major axis and other ``off-center'' positions
(\citealt{saulder23a}; Douglass et~al. 2023, in prep.).  Together with the
central spectra, these data will help constrain the total (dynamical) masses and
physical conditions in a sample of tens of thousands of \shortatlas{} galaxies.

The \shortatlas{} also has the potential to support the growing number of
time-domain and multi-messenger astronomical discoveries, wherein observations
of transient astrophysical events are detected by one or more \textit{messenger
  particles} \citep[electromagnetic radiation, neutrinos, cosmic rays, and
  gravitational waves; e.g.,][]{neronov21a}. In the case of gravitational wave
events, for example, identifying the host galaxy and, ideally, the
electromagnetic counterparts of these events is extremely challenging due to the
significant positional error ellipse of gravitational wave observations,
$\gtrsim100$~deg$^{2}$ \citep{gehrels16a, abbott20a}. Statistically complete
catalogs of large, nearby galaxies with accurate coordinates, size information,
and multi-band photometry like the \shortatlas{} are needed to help identify the
most likely source of gravitational wave events \citep[e.g.,][]{gehrels16a,
  ducoin20a, kovlakas21a}.

Finally, we highlight one other area where the \shortatlas{} is playing an
important ancillarly role. The \shortatlas{} is being used as a foreground
\textit{angular mask} for all the DESI dark-time cosmological tracers: luminous
red galaxies \citep[LRGs;][]{zhou23a}; emission-line galaxies
\citep[ELGs;][]{raichoor23a}; and quasars \citep[QSOs;][]{chaussidon23a}. Like
bright stars, large angular-diameter galaxies can bias the small-scale
clustering signal in cosmological analyses because photometric pipelines tend to
\textit{shred} structurally resolved galaxies into many smaller
sources. Similarly, the same angular mask can be used as an external input when
building photometric catalogs from other imaging data; for example, an early
version of the \shortatlas{} was used to maximize the purity of the
$3-5~\micron$ unWISE photometric catalog of two billion infrared sources
\citep{schlafly19a}. In other words, by providing a high-quality geometric mask,
the \shortatlas{} is helping these and other observational programs fulfill
their scientific promise.

\section{Summary \& Future Work}\label{sec:summary}

We present the 2020 version of the \atlas, \shortatlas, a multi-wavelength
optical and infrared imaging atlas of 383,620 large angular-diameter galaxies
covering $\approx20,000$~deg$^{2}$.\footnote{Excluding the highest regions of
dust extinction in the Galactic plane, $|b|>20\arcdeg$, the \shortatlas{} covers
74\% of the available $\approx27,200$~deg$^{2}$ of extragalactic sky.} The \shortatlas{} contains precise coordinates; optical ($grz$) and WISE/W1 through W4
($3.4-22~\micron$) infrared mosaics; model images and photometry based on the
state-of-the-art image modeling code, \thetractor; azimuthally averaged $grz$
surface-brightness and color profiles; elliptical curves of growth and
half-light radii; and extensive ancillary information for the full sample. The
complete \shortatlas{} can be accessed through the dedicated
web-portal, \url{https://sga.legacysurvey.org}.

Our measurements of the central coordinates, isophotal diameters, ellipticities,
and position angles show overall good agreement with published measurements
collated in the \leda{} extragalactic database and in the WISE Extended Source
Catalog of the 100 Largest
Galaxies \citep[WXSC-100;][]{jarrett19a}. Disagreements in these quantities
between the \shortatlas{} and \leda{} are generally due to erroneous
measurements in \leda; however, the comparison with WXSC-100 identifies small
but notable ($\approx$few arcsecond) errors in the \shortatlas{} coordinates for
some of the largest ($>5\arcmin$) galaxies in the sample.

We evaluate the completeness of the \shortatlas{} by comparing against the
HECATE all-sky, value-added galaxy catalog \citep{kovlakas21a}. We find that
the \shortatlas{} is missing approximately $1\%$ of known galaxies larger than
one arcminute and $\approx5\%$ of galaxies between $25\arcsec$ and $1\arcmin$;
however, we compute these statistics only after determining (through visual
inspection) that nearly $30\%$ of the sources larger than $1\arcmin$ in HECATE
which are missing from the \shortatlas{} are either spurious or photometric
shreds. Overall, we estimate that the \shortatlas{} is more than 95\% complete
for galaxies larger than $R(26)\approx25\arcsec$ and $r<18$ measured at the
$26$~mag~arcsec$^{-2}$ isophote in the $r$-band.

We discuss some of the potential scientific applications of the \shortatlas{}
and highlight the ongoing amassing of high-quality optical spectrophotometry
from the Dark Energy Spectroscopic
Instrument \citep[DESI;][]{desi-collaboration23b} survey. Approximately 70\% of
the \shortatlas{} sample lies within the $14,000$~deg$^{2}$ DESI footprint and
is being targeted as part of the Bright Galaxy
Survey \citep[BGS;][]{hahn23a}. To date, $7700$ DESI spectra of \shortatlas{}
galaxies are publicly available as part of the DESI Early Data
Release \citep[EDR;][]{desi-collaboration23a}, representing less than 3\% of the
final sample. In addition, the DESI Peculiar Velocity secondary
program \citep{saulder23a} is obtaining tens of thousands of major-axis,
minor-axis, and other off-center spectra of \shortatlas{} galaxies in order to
constrain the peculiar-velocity field at $z<0.15$, further enhancing the
scientific impact of the \shortatlas. Finally, we highlight the potential impact
of the \shortatlas\ for identifying the electromagnetic counterparts of
time-domain and multi-messenger astronomical events.

We conclude by discussing some of the planned improvements of the \atlas. Future
versions will focus on four broad areas, listed here in no particular order:

\begin{itemize}

\item \underline{Completeness}---As discussed
in \S\ref{sec:completeness}, the \shortatlas{} inherits the 
heterogeneity and incompleteness of the parent \leda{} sample. To mitigate these
issues, we intend to build a new parent sample of candidate large
angular-diameter galaxies by detecting them from the \shortdlis{} imaging data
themselves using a combination of traditional source-detection techniques (e.g.,
convolution kernels optimized for detecting large galaxies) and one or more
state-of-the-art deep-learning techniques \citep[e.g.,][]{stein22a,
zaritsky23a}. By inserting artificially generated galaxies of varying size,
integrated flux, and surface-brightness into the data, we intend to quantify the
completeness limits of the sample as a function of these observational
quantities.

\item \underline{Centroiding and masking}---Another area of
improvement is how the SGA pipeline handles mergers and other systems with two
or more close companions (e.g., in dense cluster environments like the Coma
Cluster), as well as galaxies near bright stars. Evaluating and optimizing the
performance of the pipeline in these circumstances will be part of a larger
visual inspection effort \citep[e.g.,][]{walmsley22a} to ensure that the central
coordinates of the largest galaxies in the \shortatlas{} are accurate (or at least
defensible, in the case of galaxies with irregular morphologies) compared with
previous independent measurements (e.g., see the discussion
in \S\ref{sec:quicksummary}).

\item \underline{Background-subtraction}---As discussed in \S\ref{sec:mosaics},
aggressive subtraction of the Mosaic-3 pattern-noise significantly impacts the
$z$-band photometry of the \shortatlas{} galaxies in the northern (BASS+MzLS)
portion of the footprint, $\mathrm{Dec}\gtrsim32\arcdeg$, including some of the
most famous and largest angular-diameter galaxies in the sample like
NGC5194=M51a and NGC5457=M101. Moreover, we show in \S\ref{sec:quicksummary}
that the outer envelopes of the early-type, spheroidal galaxies in the atlas
have likely been over-subtracted, thereby biasing their surface-brightness
profiles and inferred isophotal diameters. We intend to address both these
issues in a future version of the SGA (see, for example, \citealt{li22a} for a
new sky-subtraction technique which addresses the latter issue).  

\item \underline{Extensions in footprint and wavelength}---Since
the \shortatlas{} was finalized, thousands of square degrees of additional DECam
imaging have been acquired; we intend to use these data to extend the SGA
footprint.\footnote{See, e.g., \url{https://www.legacysurvey.org/dr10}.} In
addition, in the next version of the SGA we intend to (1) include DECam $i$-band
imaging, which is available for more than 15,000~deg$^{2}$ of area; (2)
regenerate the \unwise{} coadds but also add ultraviolet (1528~\AA{} and
2271~\AA) coadds from the Galaxy Evolution Explorer \citep[GALEX;][]{martin05a,
morrissey05a}, where GALEX data are available; and (3) measure the
surface-brightness profiles in all the availablel bandpasses from $0.15$ to
$22$~\micron. These additional measurements will further increase the scientific
impact and long-term legacy value of the SGA. 

\end{itemize}

Finally, future version of the SGA will also include ancillary spectroscopic
redshifts and spectrophotometric measurements from DESI and, eventually, a wide
range of physical properties (stellar masses, star-formation rates, etc.)
derived using state-of-the-art spectral energy distribution modeling
(Moustakas et~al. 2023, in prep.; \citealt{leja17a}).\footnote{\url{https://fastspecfit.readthedocs.io/en/latest}}

\section*{Acknowledgments}

We gratefully acknowledge specific contributions and thoughtful feedback from
Michael Blanton, Yao-Yuan Mao, and Kevin Napier, and from former Siena College
undergraduate students Alissa Ronca and Luis Villa, who contributed to an early
version of the SGA web-application. In addition, we thank Mara Salvato for
helpful comments on a draft of the paper.

JM gratefully acknowledges funding support from the U.S. Department of Energy,
Office of Science, Office of High Energy Physics under Award Number DE-SC0020086
and from the National Science Foundation under grant AST-1616414. ADM was
supported by the U.S.\ Department of Energy, Office of Science, Office of High
Energy Physics, under Award Number DE-SC0019022. AD, SJ, and BAW's research
activities are supported by the NSF's NOIRLab, which is managed by the
Association of Universities for Research in Astronomy (AURA) under a cooperative
agreement with the National Science Foundation.

We acknowledge the use of the HyperLeda database and we are especially grateful
for the time and expertise contributed by Dmitry Makarov to this project. This
research has also made extensive use of the NASA/IPAC Extragalactic Database
(NED), which is funded by the National Aeronautics and Space Administration and
operated by the California Institute of Technology; NASA’s Astrophysics Data
System; and the arXiv preprint server.

The Legacy Surveys consist of three individual and complementary projects: the
Dark Energy Camera Legacy Survey (DECaLS; Proposal ID \#2014B-0404; PIs: David
Schlegel and Arjun Dey), the Beijing-Arizona Sky Survey (BASS; NOAO Prop. ID
\#2015A-0801; PIs: Zhou Xu and Xiaohui Fan), and the Mayall z-band Legacy Survey
(MzLS; Prop. ID \#2016A-0453; PI: Arjun Dey). DECaLS, BASS and MzLS together
include data obtained, respectively, at the Blanco telescope, Cerro Tololo
Inter-American Observatory, NSF’s NOIRLab; the Bok telescope, Steward
Observatory, University of Arizona; and the Mayall telescope, Kitt Peak National
Observatory, NOIRLab. The Legacy Surveys project is honored to be permitted to
conduct astronomical research on Iolkam Du’ag (Kitt Peak), a mountain with
particular significance to the Tohono O’odham Nation.

NOIRLab is operated by the Association of Universities for Research in Astronomy
(AURA) under a cooperative agreement with the National Science Foundation.

This project used data obtained with the Dark Energy Camera (DECam), which was
constructed by the Dark Energy Survey (DES) collaboration. Funding for the DES
Projects has been provided by the U.S. Department of Energy, the U.S. National
Science Foundation, the Ministry of Science and Education of Spain, the Science
and Technology Facilities Council of the United Kingdom, the Higher Education
Funding Council for England, the National Center for Supercomputing Applications
at the University of Illinois at Urbana-Champaign, the Kavli Institute of
Cosmological Physics at the University of Chicago, Center for Cosmology and
Astro-Particle Physics at the Ohio State University, the Mitchell Institute for
Fundamental Physics and Astronomy at Texas A\&M University, Financiadora de
Estudos e Projetos, Fundacao Carlos Chagas Filho de Amparo, Financiadora de
Estudos e Projetos, Fundacao Carlos Chagas Filho de Amparo a Pesquisa do Estado
do Rio de Janeiro, Conselho Nacional de Desenvolvimento Cientifico e Tecnologico
and the Ministerio da Ciencia, Tecnologia e Inovacao, the Deutsche
Forschungsgemeinschaft and the Collaborating Institutions in the Dark Energy
Survey. The Collaborating Institutions are Argonne National Laboratory, the
University of California at Santa Cruz, the University of Cambridge, Centro de
Investigaciones Energeticas, Medioambientales y Tecnologicas-Madrid, the
University of Chicago, University College London, the DES-Brazil Consortium, the
University of Edinburgh, the Eidgenossische Technische Hochschule (ETH) Zurich,
Fermi National Accelerator Laboratory, the University of Illinois at
Urbana-Champaign, the Institut de Ciencies de l’Espai (IEEC/CSIC), the Institut
de Fisica d’Altes Energies, Lawrence Berkeley National Laboratory, the Ludwig
Maximilians Universitat Munchen and the associated Excellence Cluster Universe,
the University of Michigan, NSF’s NOIRLab, the University of Nottingham, the
Ohio State University, the University of Pennsylvania, the University of
Portsmouth, SLAC National Accelerator Laboratory, Stanford University, the
University of Sussex, and Texas A\&M University.

BASS is a key project of the Telescope Access Program (TAP), which has been
funded by the National Astronomical Observatories of China, the Chinese Academy
of Sciences (the Strategic Priority Research Program “The Emergence of
Cosmological Structures” Grant \# XDB09000000), and the Special Fund for
Astronomy from the Ministry of Finance. The BASS is also supported by the
External Cooperation Program of Chinese Academy of Sciences (Grant \#
114A11KYSB20160057), and Chinese National Natural Science Foundation (Grant \#
11433005).

The Legacy Survey team makes use of data products from the Near-Earth Object
Wide-field Infrared Survey Explorer (NEOWISE), which is a project of the Jet
Propulsion Laboratory/California Institute of Technology. NEOWISE is funded by
the National Aeronautics and Space Administration.

The Legacy Surveys imaging of the DESI footprint is supported by the Director,
Office of Science, Office of High Energy Physics of the U.S. Department of
Energy under Contract No. DE-AC02-05CH1123, by the National Energy Research
Scientific Computing Center, a DOE Office of Science User Facility under the
same contract; and by the U.S. National Science Foundation, Division of
Astronomical Sciences under Contract No. AST-0950945 to NOAO.

\software{\texttt{Astropy} \citep{astropy-collaboration13a,
astropy-collaboration18a,
astropy-collaboration22a}, \texttt{corner.py} \citep{corner}, \texttt{healpy} \citep{healpy}, \texttt{NumPy} \citep{harris20a}, \texttt{Matplotlib} \citep{hunter07a}, \texttt{Photutils} \citep{bradley23a}, \texttt{PyDL} \citep{weaver19a}, \texttt{scipy} \citep{virtanen20a}, \texttt{seaborn} \citep{waskom21a}.} 

\appendix

\section{\leda{} Database Query}\label{appendix:leda}

In \S\ref{sec:parent} we present our procedure for building the \shortatlas{}
parent sample, which begins with a \leda{} database query. For the purposes of
reproducibility, this Appendix documents the exact query we execute on the 2018
November 14 version of the \leda{} database, which results in a catalog of
1,436,176 galaxies:

\begin{verbatim}
WITH 
  "R50" AS 
  (
    SELECT pgc, avg(lax) AS lax, avg(sax) AS sax
    FROM rawdia
    WHERE quality=0 and dcode=5 and band between 4400 and 4499 GROUP BY pgc
  ),
  "IR" AS 
  (
    SELECT pgc, avg(lax) AS lax, avg(sax) AS sax
    FROM rawdia
    WHERE quality=0 and iref in (27129) and dcode=7 and band=0 GROUP BY pgc
  )
SELECT
  m.pgc, m.objname, m.objtype, m.al2000, m.de2000, m.type, 
  m.bar, m.ring, m.multiple, m.compactness, m.t, m.logd25, 
  m.logr25, m.pa, m.bt, m.it, m.kt, m.v, m.modbest, 
  "R50".lax, "R50".sax, "IR".lax, "IR".sax,
FROM
  m000 AS m
  LEFT JOIN "R50" USING (pgc)
  LEFT JOIN "IR" USING (pgc)
WHERE
  objtype='G' and (m.logd25>0.2 or "R50".lax>0.2 or "IR".lax>0.2)
\end{verbatim}

\section{Data Products \& Data Access}\label{appendix:data}

In this Appendix, we define the data model for the \shortatlas{} data products
and describe how all the data can be accessed. All the input imaging data used
to construct the \shortatlas, including the six-year \unwise{} image stacks, are
publicly accessible
through \url{https://www.legacysurvey.org}. The \shortatlas{} data themselves
can be retrieved through the \textit{SGA-2020 Data Portal} at
\url{https://sga.legacysurvey.org}. This portal provides a searchable database
for retrieving data and visualizations for individual galaxies, as well as
bulk-download access to the full dataset. For example, the following link will
navigate directory to the beautiful face-on spiral (Hubble type SABc)
NGC2532: \url{https://sga.legacysurvey.org/group/NGC2532}. In addition, the data
can be accessed (and cross-referenced against a growing number of additional
datasets, including DESI) through NSF's NOIRLab Astro Data Lab at
\url{https://datalab.noirlab.edu/sga.php}. Finally, one can interactively explore
the sample in the \shortdlis{} imaging footprint via
the \texttt{SGA} \textit{layer} of the \textit{Legacy Surveys Viewer}; for
example, the following link will navigate to the position of NGC2532 in the
North Galactic
Cap: \url{https://www.legacysurvey.org/viewer?ra=122.5638&dec=33.9568&layer=ls-dr9&zoom=13&sga}.

Most users will be interested in the \texttt{SGA-2020.fits} file, a
multi-extension FITS catalog which contains detailed information for all 383,620
galaxies in the \shortatlas. Table~\ref{table:sga2020} summarizes the contents
of this file and Table~\ref{table:sga2020-ellipse} contains a detailed
description of the \texttt{ELLIPSE} Header/Data Unit (HDU). In addition,
Table~\ref{table:bitmasks} defines the \texttt{ELLIPSEBIT} bit-masks which
record issues associated with the ellipse-fitting and surface-brightness profile
modeling, if any.

Furthermore, for each galaxy group in the atlas (i.e., each row in
the \texttt{SGA-2020.fits} catalog where \texttt{GROUP\_PRIMARY}
is \texttt{True}), we generate the set of files summarized in
Table~\ref{table:images} (see also \S\ref{sec:ellipse}), which include the
individual multi-wavelength mosaics, \tractor{} catalogs, surface-brightness
profiles, and other key data products. These files are organized into the
directory structure \texttt{RASLICE/GROUP\_NAME}, where
\texttt{GROUP\_NAME} is the name of the galaxy group (see \S\ref{sec:groups})
and \texttt{RASLICE}, which ranges between \texttt{000} and \texttt{359}, is the
one-degree wide slice of the sky that the object belongs to. Specifically, in
Python:
\begin{verbatim}
  RASLICE = "{:03d}".format(GROUP_RA)
\end{verbatim}

\noindent Finally, Table~\ref{table:ellipse} documents the data model of the
ellipse-fitting and surface-brightness profile results for each individual
galaxy in the \shortatlas.

\section{Known Issues}\label{appendix:issues}

Not surprisingly, a catalog the size and complexity of the \shortatlas{} has
imperfections, most of which were identified after the fitting was finalized. In
this Appendix we document the currently known issues, all of which we intend to
to address in future versions of the SGA. For additional details and the most
up-to-date documentation regarding these and other issues, we refer the
interested reader to \url{https://github.com/moustakas/SGA/issues}.

In terms of the data reduction, the most significant known issue impacting
the \shortatlas{} is the pattern-noise subtraction from the Mosaic-3 imaging,
which distorts the $z$-band surface-brightness profiles and colors of galaxies
in the (northern) BASS+MzLS portion of the footprint (see \S~\ref{sec:mosaics}
for details). In addition, SGA users have reported that in some cases the WISE
W1 and W2 background signal has been over-subtracted. We hypothesize that this
over-subtraction arises because the unWISE background is modeled as the median
flux after subtracting all the point-sources in a $1\arcmin\times1\arcmin$
grid \citep{schlafly19a}, which would potentially impact all galaxies in the
sample larger than approximately one arcminute. 

Some issues also impact the availability and accuracy of the optical photometry
in the \shortatlas. The most significant problem is that the elliptical aperture
photometry (and the corresponding curves of growth) reported in the
individual \texttt{GROUP\_NAME}-largegalaxy-\texttt{SGA\_ID}-ellipse.fits files
(see Table~\ref{table:images}) were impacted by a catastrophic bug which
rendered these measurements unusable. The data model for this file is documented
in Table~\ref{table:ellipse} and the impacted columns are flagged with the text
``Do not use; see Appendix~\ref{appendix:issues}.'' Fortunately, however, we
were able to recover the curves of growth from the surface-brightness profiles
themselves, which were not affected by this bug; we record those measurements in
the \texttt{ELLIPSE} HDU of the merged \texttt{SGA-2020.fits} file (as
documented in Table~\ref{table:sga2020-ellipse}).

Moreover, ellipse-fitting was skipped, failed, or rejected for a few different
reasons which are encoded in the \texttt{ELLIPSEBIT} column; this column appears
in Table~\ref{table:sga2020-ellipse} and is documented in
Table~\ref{table:bitmasks}. First, a total of 8415 galaxies were not ellipse-fit
because they were deemed to be too small
($\texttt{ELLIPSEBIT}=2^{1}\, \texttt{or}\, \texttt{ELLIPSEBIT}=2^{2}$) via
their best-fitting \tractor{} model and half-light radius (\texttt{shape\_r});
specifically, objects modeled by \thetractor{} as \texttt{type=REX} with
$\texttt{shape\_r}<2\arcsec$ or \texttt{type=\{EXP,DEV,SER\}} with
$\texttt{shape\_r}<5\arcsec$ were not
ellipse-fit.\footnote{\thetractor{} \texttt{type} and \texttt{shape\_r}
measurements are fully documented
at \url{https://www.legacysurvey.org/dr9/catalogs/\#region-tractor-aaa-tractor-brick-fits}.}
Second, ellipse-fitting did not complete ($\texttt{ELLIPSEBIT}=2^{4}$) for 27
galaxies in the following seven groups: IC1613, NGC0055 Group, NGC0253 Group,
NGC0300 Group, NGC0598 Group, NGC3031 Group, and NGC5457. The central galaxies
in these groups rank among the largest in the sample
($18\arcmin<\texttt{GROUP\_DIAMETER}<62\arcmin$), so modeling these systems
posed some especially acute computational challenges. Third, the ellipse-fitting
results for 52 galaxies were rejected ($\texttt{ELLIPSEBIT}=2^{5}$) because they
were found via visual inspection to be incorrect or unreliable, usually due to
incomplete or imperfect masking of nearby bright stars or other
galaxies. Finally, ellipse-fitting was not carried out (or silently failed) on
6161 galaxies where the pipeline did not indicate a problem and therefore
$\texttt{ELLIPSEBIT}=0$ for these systems even though there are no measured
surface-brightness profiles.

In no particular order, some of the additional currently known issues include:
\begin{itemize}
\item A small number of objects without imaging in all three $grz$ bandpasses
(e.g., \url{https://sga.legacysurvey.org/group/PGC1411139}) made it into final
sample, despite the requirement of three-band optical imaging discussed
in \S\ref{sec:ellipse};
\item A small fraction of objects have compromised surface-brightness profiles
due to their proximity to bright stars
(e.g., \url{https://sga.legacysurvey.org/group/UGC09630}) or because of poor
deblending of multiple sources (typically due to incompleteness in the
parent \leda{} catalog;
e.g., \url{https://sga.legacysurvey.org/group/PGC087438_GROUP}).  
\item The central coordinates for a small number of the largest ($>5\arcmin$)
galaxies in the \shortatlas{} are incorrect by up to a few arcseconds, as
discussed in \S\ref{sec:quicksummary}; 
\item In a handful of spheroidal galaxies with large ellipticity ($\epsilon>0.5$
or $b/a<0.5$), our masking algorithm (see \S~\ref{sec:tractor}) is too
aggressive and inadvertently masks some of the outer-envelope light of the
galaxy along the minor axis
(e.g., \url{https://sga.legacysurvey.org/group/IC0941}). 
\end{itemize}

\newpage

\begin{deluxetable}{ccll}
\tabletypesize{\footnotesize}
\tablecaption{\texttt{SGA-2020.fits} File Contents \label{table:sga2020}}
\tablewidth{0pt}
\tablehead{
\multicolumn{2}{c}{Extension} & 
\colhead{Data Model} &
\colhead{} \\
\colhead{Number} & 
\colhead{Name} & 
\colhead{Documentation} &
\colhead{Description}
}
\startdata
HDU01 & \texttt{ELLIPSE} & Table~\ref{table:sga2020-ellipse} & Sample metadata and ellipse-fitting results. \\
HDU02 & \texttt{TRACTOR} & LS/DR9 Website\tablenotemark{\footnotesize{a}} & \thetractor{} fitting results.\tablenotemark{\footnotesize{b}}
\enddata
\tablecomments{Catalog containing the principal measurements for all 383,620 galaxies in the \shortatlas{} as a
multi-extension FITS file with two row-matched Header/Data Units (HDUs).}
\tablenotetext{a}{\url{https://www.legacysurvey.org/dr9/catalogs}}
\tablenotetext{b}{This table also includes \texttt{SGA\_ID} (see
Table~\ref{table:sga2020-ellipse}) to facilitate cross-matching.}
\end{deluxetable}

\startlongtable
\begin{deluxetable}{lcl}
\tabletypesize{\scriptsize}
\tablecaption{\texttt{ELLIPSE} HDU Data Model\label{table:sga2020-ellipse}}
\tablewidth{0pt}
\tablehead{
\colhead{Column Name} & 
\colhead{Units} & 
\colhead{Description} 
}
\startdata
\texttt{SGA\_ID}                                                   &  \nodata                                     & Unique integer identifier.  \\
\texttt{SGA\_GALAXY}                                               &  \nodata                                     & SGA galaxy name, constructed as ``SGA-2020-\texttt{SGA\_ID}''.  \\
\texttt{GALAXY}                                                    &  \nodata                                     & Unique galaxy name.  \\
\texttt{PGC}\tablenotemark{\scriptsize{a}}                         &  \nodata                                     & Unique identifier from the \textit{Principal Catalogue of Galaxies}.  \\
\texttt{RA\_LEDA}                                                  &  degree                                      & Right ascension (J2000) from the reference indicated in \texttt{REF}. \\
\texttt{DEC\_LEDA}                                                 &  degree                                      & Declination (J2000) from the reference indicated in \texttt{REF}. \\
\texttt{MORPHTYPE}                                                 &  \nodata                                     & Visual morphological type from \leda{} (if available).  \\
\texttt{PA\_LEDA}                                                  &  degree                                      & Galaxy position angle (measured East of North); taken from \\
                                                                   &                                              & the reference indicated in \texttt{REF}.  \\
\texttt{D25\_LEDA}                                                 &  arcmin                                      & Major-axis diameter at the $25$~mag~arcsec$^{-2}$ (optical) surface  \\
                                                                   &                                              & brightness isophote; taken from the reference indicated in \texttt{REF}. \\
\texttt{BA\_LEDA}                                                  &  \nodata                                     & Ratio of the semi-minor axis to the semi-major axis; taken from  \\
                                                                   &                                              & the reference indicated in \texttt{REF}. \\
\texttt{Z\_LEDA}\tablenotemark{\scriptsize{a}}                     &  \nodata                                     & Heliocentric redshift from \leda. \\ 
\texttt{SB\_D25\_LEDA}                                             &  Vega~mag~arcsec$^{-2}$                       & Mean surface brightness based on \texttt{D25\_LEDA} and \texttt{MAG\_LEDA}.  \\
\texttt{MAG\_LEDA}\tablenotemark{\scriptsize{b}}                   &  Vega mag                                    & Apparent $b_{t}$-band magnitude. \\ 
\texttt{BYHAND}                                                    &  \nodata                                     & Boolean  flag indicating whether one or more of \texttt{RA\_LEDA}, \texttt{DEC\_LEDA}, \\
                                                                   &                                              & \texttt{D25\_LEDA}, \texttt{PA\_LEDA}, \texttt{BA\_LEDA}, or \texttt{MAG\_LEDA} were changed from their  \\
                                                                   &                                              & published values (usually via visual inspection) while building the \\
                                                                   &                                              & parent sample.  \\
\texttt{REF}                                                       & \nodata                                      & Reference indicating the origin of the object (see \S\ref{sec:sample}): \\
                                                                   &                                              & \texttt{LEDA-20181114}, \texttt{LGDWARFS}, \texttt{RC3}, \texttt{OpenNGC}, or \texttt{DR8}.  \\
\texttt{GROUP\_ID}                                                 & \nodata                                      & Unique group identification number.  \\
\texttt{GROUP\_NAME}                                               & \nodata                                      & Group name, constructed from the name of its largest member   \\
                                                                   &                                              & (see \S\ref{sec:groups}). For isolated galaxies, identical to \texttt{GALAXY}. \\
\texttt{GROUP\_MULT}                                               & \nodata                                      & Number of group members (i.e., group multiplicity).  \\
\texttt{GROUP\_PRIMARY}                                            & \nodata                                      & Boolean flag indicating the primary (i.e., largest) group member.  \\
\texttt{GROUP\_RA}                                                 & degree                                       & Mean right ascension of the group weighted by \texttt{D25\_LEDA}.  \\
\texttt{GROUP\_DEC}                                                & degree                                       & Mean declination of the group weighted by \texttt{D25\_LEDA}.  \\
\texttt{GROUP\_DIAMETER}                                           & arcmin                                       & Approximate group diameter (see \S\ref{sec:groups}). \\
\texttt{BRICKNAME}                                                 & \nodata                                      & Name of custom \tractor{} ``brick'', encoding the sky position, e.g. \\
                                                                   &                                              & ``1126p222'' is centered on RA=112.6, Dec=+22.2.  \\ 
\texttt{RA}                                                        & degree                                       & Right ascension (J2000) based on \thetractor{} model fit. \\
\texttt{DEC}                                                       & degree                                       & Declination (J2000) based on \thetractor{} model fit. \\ 
\texttt{D26}                                                       & arcmin                                       & Major axis diameter at the $\mu=26$~mag~arcsec$^{-2}$ $r$-band isophote. \\
\texttt{D26\_REF}\tablenotemark{\scriptsize{c}}                    & \nodata                                      & Reference indicating the origin of the \texttt{DIAM} measurement: \\
                                                                   &                                              & \texttt{SB26}, \texttt{SB25}, or \texttt{LEDA}.  \\
\texttt{PA}                                                        & degree                                       & Galaxy position angle (measured East of North), as  \\
                                                                   &                                              & measured from the \texttt{ellipse moments} (see \S\ref{sec:sbprofiles}) (or equivalent to \\
                                                                   &                                              & \texttt{PA\_LEDA} if the \texttt{ellipse moments} could not be measured).  \\
\texttt{BA}                                                        & \nodata                                      & Minor-to-major axis ratio, as measured from the \texttt{ellipse moments} (or \\
                                                                   &                                              & equivalent to \texttt{BA\_LEDA} if the \texttt{ellipse moments} could not be measured).  \\
\texttt{RA\_MOMENT}                                                & degree                                       & Light-weighted right ascension (J2000), as measured from the \texttt{ellipse} \\
                                                                   &                                              & \texttt{moments}. Equivalent to \texttt{RA\_X0} in Table~\ref{table:ellipse} but set to \texttt{RA\_LEDA} if \\
                                                                   &                                              & ellipse-fitting failed or was not carried out.  \\
\texttt{DEC\_MOMENT}                                               & degree                                       & Like \texttt{RA\_MOMENT} but for the declination axis. \\
\texttt{SMA\_MOMENT}\tablenotemark{\scriptsize{d}}                 & arcsec                                       & Second moment of the light distribution along the major axis based    \\
                                                                   &                                              & on the measured \texttt{ellipse moments}. Equivalent to \texttt{MAJORAXIS} in \\ 
                                                                   &                                              & Table~\ref{table:ellipse} but converted to arcsec.  \\
\texttt{\textlangle grz\textrangle\_SMA50}\tablenotemark{\scriptsize{d}} & arcsec                                 & Half-light semi-major axis length based on equation~(\ref{eq:halflight}). \\
\texttt{SMA\_SB\textlangle sblevel\textrangle}\tablenotemark{\scriptsize{d}} & arcsec                             & Semi-major axis length at the $r$-band \texttt{\textlangle sblevel\textrangle} mag~arcsec$^{-2}$ isophote. \\
\texttt{\textlangle grz\textrangle\_MAG\_SB\textlangle sblevel\textrangle}\tablenotemark{\scriptsize{d}} & AB mag & Cumulative brightness measured within \texttt{SMA\_SB\textlangle sblevel\textrangle}.  \\
\texttt{\textlangle grz\textrangle\_MAG\_SB\textlangle sblevel\textrangle\_ERR}\tablenotemark{\scriptsize{d}} & AB mag & $1\sigma$ uncertainty in \texttt{\textlangle grz\textrangle\_MAG\_SB\textlangle sblevel\textrangle}. \\
\texttt{\textlangle grz\textrangle\_COG\_PARAMS\_MTOT}\tablenotemark{\scriptsize{d}}   & AB mag                   & Best-fitting curve-of-growth parameter $m_{1}$ from equation~(\ref{eq:curveofgrowth}). \\
\texttt{\textlangle grz\textrangle\_COG\_PARAMS\_M0}\tablenotemark{\scriptsize{d}}     & AB mag                   & Best-fitting curve-of-growth parameter $m_{0}$ from equation~(\ref{eq:curveofgrowth}). \\
\texttt{\textlangle grz\textrangle\_COG\_PARAMS\_ALPHA1}\tablenotemark{\scriptsize{d}} & \nodata                  & Best-fitting curve-of-growth parameter $\alpha_{1}$ from equation~(\ref{eq:curveofgrowth}). \\
\texttt{\textlangle grz\textrangle\_COG\_PARAMS\_ALPHA2}\tablenotemark{\scriptsize{d}} & \nodata                  & Best-fitting curve-of-growth parameter $\alpha_{2}$ from equation~(\ref{eq:curveofgrowth}). \\
\texttt{\textlangle grz\textrangle\_COG\_PARAMS\_CHI2}\tablenotemark{\scriptsize{d}}   & \nodata                  & $\chi^{2}$ of the fit to the curve-of-growth (see \S\ref{sec:sbprofiles}).  \\
\texttt{ELLIPSEBIT}                                                & \nodata                                      & See Table~\ref{table:bitmasks}. \\
\enddata
\tablecomments{\texttt{ELLIPSE} HDU of the \texttt{SGA-2020.fits} merged catalog defined in Table~\ref{table:sga2020}. 
In this table \texttt{\textlangle grz\textrangle} denotes the $g$-, $r$-, or
$z$-band filter; \texttt{\textlangle wise\textrangle} denotes the $W1$-, $W2$-,
$W3$-, or $W4$ bandpass; and \texttt{\textlangle sblevel\textrangle} represents
the 22, 22.5, 23, 23.5, 24, 24.5, 25, 25.5, and 26~mag~arcsec$^{-2}$ isophote.}
\tablenotetext{a}{Missing values are represented with a $-1$. For some
  quantities (e.g., \texttt{PGC} or \texttt{Z\_LEDA}) a missing value does not
  necessarily mean that that value does not exist.}
\tablenotetext{b}{This magnitude estimate is heterogeneous in both bandpass and
  aperture but for most galaxies it is measured in the $B$-band; use with care.}
\tablenotetext{c}{By default, we infer \texttt{D26} from
  \texttt{SMA\_SB26}. However, if the $r$-band surface-brightness profile could
  not be measured at this level, we estimate \texttt{D26} as
  $2.5\times$\texttt{SMA\_SB25} or $1.5\times$\texttt{D25\_LEDA}, in that order
  of priority.}
\tablenotetext{d}{If ellipse-fitting or curve-of-growth modeling failed or was not attempted then these columns' values are $-1$.}
\end{deluxetable}

\newpage

\begin{deluxetable}{cll}
\tabletypesize{\footnotesize}
\tablecaption{Ellipse-Fitting Bitmasks \label{table:bitmasks}}
\tablewidth{0pt}
\tablehead{
\colhead{Bit} & 
\colhead{Bit} & 
\colhead{}  \\
\colhead{Number} & 
\colhead{Name} & 
\colhead{Definition} 
}
\startdata
0 & \nodata                   & Not used; ignore. \\
1 & \texttt{REX\_TOOSMALL}    & Ellipse-fit skipped; galaxy classified as too-small type \texttt{REX}. \\
2 & \texttt{NOTREX\_TOOSMALL} & Ellipse-fit skipped; galaxy classified as too-small type \texttt{EXP}, \texttt{DEV}, or \texttt{SER}. \\
3 & \texttt{FAILED}           & Ellipse-fitting was attempted but failed. \\
4 & \texttt{NOTFIT}           & Ellipse-fitting was not attempted. \\
5 & \texttt{REJECTED}         & Ellipse-fitting results were rejected based on visual inspection. \\
\enddata
\tablecomments{Bitmask encoding various reasons why ellipse-fitting failed or was not attempted; see \S\ref{sec:sbprofiles} for additional details.} 
\end{deluxetable}

\begin{deluxetable}{ll}
\tabletypesize{\scriptsize}
\tablecaption{Images and Catalogs\label{table:images}}
\tablewidth{0pt}
\tablehead{
\colhead{Filename} & 
\colhead{Description} 
}
\startdata
\texttt{GROUP\_NAME}-ccds-\texttt{\textlangle region\textrangle}.fits               & CCDs contributing to the optical image stacks. \\
\texttt{GROUP\_NAME}-largegalaxy-blobs.fits.gz                                      & Enumerated segmentation (``blob'') image. \\
\texttt{GROUP\_NAME}-largegalaxy-tractor.fits                                       & \tractor{} catalog of all detected sources in the field. \\
\texttt{GROUP\_NAME}-largegalaxy-maskbits.fits.fz                                   & Image encoding the LS/DR9 bitmasks\tablenotemark{\scriptsize{a}} contributing to each pixel. \\
\texttt{GROUP\_NAME}-largegalaxy-outlier-mask.fits.fz                               & Image of pixels rejected during outlier masking. \\
\texttt{GROUP\_NAME}-depth-\texttt{\textlangle grz\textrangle}.fits.fz              & Image of the $5\sigma$ point-source depth at each pixel. \\
\texttt{GROUP\_NAME}-largegalaxy-psf-\texttt{\textlangle grz\textrangle}.fits.fz    & Postage stamp of the inverse-variance weighted mean pixelized \\
                                                                                    & PSF at the center of the field.  \\
\texttt{GROUP\_NAME}-largegalaxy-image-\texttt{\textlangle grz\textrangle}.fits.fz  & Inverse-variance weighted optical image. \\
\texttt{GROUP\_NAME}-largegalaxy-invvar-\texttt{\textlangle grz\textrangle}.fits.fz & Optical inverse variance image stack. \\
\texttt{GROUP\_NAME}-largegalaxy-model-\texttt{\textlangle grz\textrangle}.fits.fz  & Optical \tractor{} model image coadd. \\
\texttt{GROUP\_NAME}-largegalaxy-image-grz.jpg                                      & Color ($grz$) \textsc{jpg} image of the image stack. \\
\texttt{GROUP\_NAME}-largegalaxy-model-grz.jpg                                      & Color ($grz$) \textsc{jpg} image of \thetractor{} model image. \\
\texttt{GROUP\_NAME}-largegalaxy-resid-grz.jpg                                      & Color ($grz$) \textsc{jpg} image of residual (data minus model) image. \\
\texttt{GROUP\_NAME}-image-\texttt{\textlangle wise\textrangle}.fits.fz             & Inverse-variance weighted infrared image stack. \\
\texttt{GROUP\_NAME}-invvar-\texttt{\textlangle wise\textrangle}.fits.fz            & Infrared inverse variance image stack. \\
\texttt{GROUP\_NAME}-largegalaxy-model-\texttt{\textlangle wise\textrangle}.fits.fz & Infrared \tractor{} model image coadd.  \\
\texttt{GROUP\_NAME}-image-W1W2.jpg                                                 & Color ($W1W2$) \textsc{jpg} image of the image stack. \\
\texttt{GROUP\_NAME}-model-W1W2.jpg                                                 & Color ($W1W2$) \textsc{jpg} image of \thetractor{} model image. \\
\texttt{GROUP\_NAME}-largegalaxy-sample.fits                                        & Catalog of one or more galaxies from the parent sample in this group. \\
\texttt{GROUP\_NAME}-largegalaxy-\texttt{SGA\_ID}-ellipse.fits\tablenotemark{\scriptsize{b}} & Detailed ellipse-fitting results for each galaxy in this group; see \\
                                                                                    & Table~\ref{table:ellipse} for the data model. \\
\texttt{GROUP\_NAME}-coadds.log                                                     & Log output for the \texttt{coadds} stage of the SGA pipeline. \\
\texttt{GROUP\_NAME}-ellipse.log                                                    & Log output for the \texttt{ellipse} stage of the SGA pipeline. \\
\enddata
\tablecomments{Summary of files generated for each galaxy group in the \shortatlas. In this table, \texttt{\textlangle region\textrangle} denotes either \texttt{north} for BASS/MzLS or
\texttt{south} for DECaLS; \texttt{\texttt{\textlangle grz\textrangle}} denotes the $g$-, $r$-, or $z$-band filter; 
and \texttt{\textlangle wise\textrangle} denotes the $W1$-, $W2$-, $W3$-, or
$W4$ bandpass.}
\tablenotetext{a}{\url{https://www.legacysurvey.org/dr9/bitmasks}}
\tablenotetext{b}{This file may be missing (for the galaxy of a given \texttt{SGA\_ID}) if ellipse-fitting fails or is not carried out.}  
\end{deluxetable}

\startlongtable
\begin{deluxetable}{lcl}
\tabletypesize{\scriptsize}
\tablecaption{Ellipse-Fitting Data Model\label{table:ellipse}}
\tablewidth{0pt}
\tablehead{
\colhead{Column Name} & 
\colhead{Units} & 
\colhead{Description} 
}
\startdata
\texttt{SGA\_ID}                                               &  \nodata                  & See Table~\ref{table:sga2020-ellipse}. \\
\texttt{GALAXY}                                                &  \nodata                  & See Table~\ref{table:sga2020-ellipse}. \\
\texttt{RA}                                                    &  degree                   & See Table~\ref{table:sga2020-ellipse}. \\
\texttt{DEC}                                                   &  degree                   & See Table~\ref{table:sga2020-ellipse}. \\
\texttt{PGC}                                                   &  \nodata                  & See Table~\ref{table:sga2020-ellipse}. \\
\texttt{PA\_LEDA}                                              &  degree                   & See Table~\ref{table:sga2020-ellipse}. \\
\texttt{BA\_LEDA}                                              &  \nodata                  & See Table~\ref{table:sga2020-ellipse}. \\
\texttt{D25\_LEDA}                                             &  arcmin                   & See Table~\ref{table:sga2020-ellipse}. \\
\texttt{BANDS}                                                 &  \nodata                  & List of bandpasses fitted (here, always $grz$). \\
\texttt{REFBAND}                                               &  \nodata                  & Reference band (here, always $r$). \\
\texttt{REFPIXSCALE}                                           &  arcsec~pixel$^{-1}$       & Pixel scale in \texttt{REFBAND}. \\
\texttt{SUCCESS}                                               &  \nodata                  & Flag indicating ellipse-fitting success or failure. \\
\texttt{FITGEOMETRY}                                           &  \nodata                  & Flag indicating whether the ellipse geometry was allowed to vary  \\
                                                               &                           & with semi-major axis (here, always \texttt{False}). \\
\texttt{INPUT\_ELLIPSE}                                        &  \nodata                  & Flag indicating whether ellipse parameters were passed from an  \\
                                                               &                           & external file (here, always \texttt{False}). \\
\texttt{LARGESHIFT}                                            &  \nodata                  & Flag indicating that the light-weighted center (from \texttt{ellipse moments}) \\
                                                               &                           & is different from \thetractor{} position by more than 10 pixels in either \\
                                                               &                           & dimension, in which case we adopt \thetractor{} model position. \\
\texttt{RA\_X0}                                                &  degree                   & Right ascension (J2000) at pixel position \texttt{X0}. \\
\texttt{DEC\_Y0}                                               &  degree                   & Declination (J2000) at pixel position \texttt{Y0}. \\
\texttt{X0}                                                    &  pixel                    & Light-weighted position along the $x$-axis (from \texttt{ellipse moments}). \\
\texttt{Y0}                                                    &  pixel                    & Light-weighted position along the $y$-axis (from \texttt{ellipse moments}). \\
\texttt{EPS}                                                   &  \nodata                  & Ellipticity, $\epsilon\equiv1-b/a$, where $b/a$ is the semi-minor to semi-major \\
                                                               &                           & axis ratio \texttt{BA} in Table~\ref{table:sga2020}. \\
\texttt{PA}                                                    &  degree                   & Galaxy position angle (astronomical convention, measured East of North\\
                                                               &                           & ); equivalent to \texttt{PA} in Table~\ref{table:sga2020}. \\
\texttt{THETA}                                                 &  degree                   & Galaxy position angle (physics convention, measured North of West) \\
                                                               &                           & given by $270-\mbox{\texttt{PA}\ \texttt{mod}}\ 180$. \\ 
\texttt{MAJORAXIS}                                             &  pixel                    & Light-weighted length of the semi-major axis (from \texttt{ellipse moments}). \\
\texttt{MAXSMA}                                                &  pixel                    & Maximum semi-major axis length used for the ellipse-fitting and  \\
                                                               &                           & curve-of-growth measurements (typically taken to be $2\times\mbox{\texttt{MAJORAXIS}}$). \\
\texttt{INTEGRMODE}\tablenotemark{\scriptsize{a}}              &  \nodata                  & Integration mode (here, always \textit{median}). \\
\texttt{SCLIP}\tablenotemark{\scriptsize{a}}                   &  \nodata                  & Sigma-clipping threshold (here, always \textit{3}). \\
\texttt{NCLIP}\tablenotemark{\scriptsize{a}}                   &  \nodata                  & Sigma-clipping iterations (here, always \textit{2}). \\
\texttt{PSFSIZE\_\textlangle grz\textrangle}                   &  arcsec                   & Mean FWHM of the point-spread function over the full mosaic (derived \\
                                                               &                           & from the \texttt{PSFSIZE\_\textlangle grz\textrangle} columns in \thetractor{} catalogs). \\
\texttt{PSFDEPTH\_\textlangle grz\textrangle}                  &  AB mag                   & Mean $5\sigma$ point-source depth over the full mosaic (derived from the \\
                                                               &                           & \texttt{PSFDEPTH\_\textlangle grz\textrangle} columns in \thetractor{} catalogs). \\
\texttt{MW\_TRANSMISSION\_\textlangle grz\textrangle}          &  \nodata                  & Galactic transmission fraction (taken from the corresponding \\
                                                               &                           & \tractor{} catalog at the central coordinates of the galaxy). \\
\texttt{REFBAND\_WIDTH}                                        &  pixel                    & Width of the optical mosaics in \texttt{REFBAND}. \\
\texttt{REFBAND\_HEIGHT}                                       &  pixel                    & Height of the optical mosaics in \texttt{REFBAND}. \\
\texttt{\textlangle grz\textrangle\_SMA}                       &  pixel                    & Ellipse semi-major axis position. \\ 
\texttt{\textlangle grz\textrangle\_INTENS}                    &  nanomaggies~arcsec$^{-2}$ & Linear surface brightness at \texttt{\textlangle grz\textrangle\_SMA}. \\
\texttt{\textlangle grz\textrangle\_INTENS\_ERR}               &  nanomaggies~arcsec$^{-2}$ & $1\sigma$ uncertainty in \texttt{\textlangle grz\textrangle\_INTENS}. \\
\texttt{\textlangle grz\textrangle\_EPS}                       &  \nodata                  & Ellipse ellipticity; here, fixed at \texttt{EPS}. \\
\texttt{\textlangle grz\textrangle\_EPS\_ERR}                  &  \nodata                  & $1\sigma$ uncertainty in \texttt{\textlangle grz\textrangle\_EPS}. \\
\texttt{\textlangle grz\textrangle\_PA}                        &  degree                   & Ellipse position angle; here, fixed at \texttt{PA}. \\
\texttt{\textlangle grz\textrangle\_PA\_ERR}                   &  degree                   & $1\sigma$ uncertainty in \texttt{\textlangle grz\textrangle\_PA}. \\
\texttt{\textlangle grz\textrangle\_X0}                        &  pixel                    & Ellipse $x$-axis pixel coordinate; here, fixed at \texttt{X0}. \\
\texttt{\textlangle grz\textrangle\_X0\_ERR}                   &  pixel                    & $1\sigma$ uncertainty in \texttt{\textlangle grz\textrangle\_X0}. \\
\texttt{\textlangle grz\textrangle\_Y0}                        &  pixel                    & Ellipse $y$-axis pixel coordinate; here, fixed at \texttt{Y0}. \\
\texttt{\textlangle grz\textrangle\_Y0\_ERR}                   &  pixel                    & $1\sigma$ uncertainty in \texttt{\textlangle grz\textrangle\_Y0}. \\                       
\texttt{\textlangle grz\textrangle\_A3}\tablenotemark{\scriptsize{b}}      &  \nodata & Third-order harmonic coefficient; not used. \\
\texttt{\textlangle grz\textrangle\_A3\_ERR}\tablenotemark{\scriptsize{b}} &  \nodata & $1\sigma$ uncertainty in \texttt{\textlangle grz\textrangle\_A3}. \\                       
\texttt{\textlangle grz\textrangle\_A4}\tablenotemark{\scriptsize{b}}      &  \nodata & Fourth-order harmonic coefficient; not used. \\
\texttt{\textlangle grz\textrangle\_A4\_ERR}\tablenotemark{\scriptsize{b}} &  \nodata & $1\sigma$ uncertainty in \texttt{\textlangle grz\textrangle\_A4}. \\
\texttt{\textlangle grz\textrangle\_RMS}\tablenotemark{\scriptsize{b}}         &  nanomaggies~arcsec$^{-2}$ & Root-mean-square surface brightness along the elliptical path. \\
\texttt{\textlangle grz\textrangle\_PIX\_STDDEV}\tablenotemark{\scriptsize{b}} &  nanomaggies              & Pixel standard deviation estimate. \\
\texttt{\textlangle grz\textrangle\_STOP\_CODE}\tablenotemark{\scriptsize{b}}  &  \nodata                  & Fitting stop code. \\
\texttt{\textlangle grz\textrangle\_NDATA}\tablenotemark{\scriptsize{b}}       &  \nodata                  & Number of data points used for the fit. \\
\texttt{\textlangle grz\textrangle\_NFLAG}\tablenotemark{\scriptsize{b}}       &  \nodata                  & Number of points rejected during the fit. \\
\texttt{\textlangle grz\textrangle\_NITER}\tablenotemark{\scriptsize{b}}       &  \nodata                  & Number of fitting iterations. \\
\texttt{\textlangle grz\textrangle\_COG\_SMA}                  & pixel                   & Do not use; see Appendix~\ref{appendix:issues}. \\
\texttt{\textlangle grz\textrangle\_COG\_MAG}                  & AB mag                  & Do not use; see Appendix~\ref{appendix:issues}. \\
\texttt{\textlangle grz\textrangle\_COG\_MAGERR}               & AB mag                  & Do not use; see Appendix~\ref{appendix:issues}. \\
\texttt{\textlangle grz\textrangle\_COG\_PARAMS\_MTOT}         & AB mag                  & Do not use; see Appendix~\ref{appendix:issues}. \\
\texttt{\textlangle grz\textrangle\_COG\_PARAMS\_M0}           & AB mag                  & Do not use; see Appendix~\ref{appendix:issues}. \\
\texttt{\textlangle grz\textrangle\_COG\_PARAMS\_ALPHA1}       & \nodata                 & Do not use; see Appendix~\ref{appendix:issues}. \\
\texttt{\textlangle grz\textrangle\_COG\_PARAMS\_ALPHA2}       & \nodata                 & Do not use; see Appendix~\ref{appendix:issues}. \\
\texttt{\textlangle grz\textrangle\_COG\_PARAMS\_CHI2}         & \nodata                 & Do not use; see Appendix~\ref{appendix:issues}. \\
\texttt{RADIUS\_SB\textlangle sblevel\textrangle}              & arcsec                  & Do not use; see Appendix~\ref{appendix:issues}. \\
\texttt{RADIUS\_SB\textlangle sblevel\textrangle\_ERR}         & arcsec                  & Do not use; see Appendix~\ref{appendix:issues}. \\
\texttt{\textlangle grz\textrangle\_MAG\_SB\textlangle sblevel\textrangle}      & AB mag & Do not use; see Appendix~\ref{appendix:issues}. \\
\texttt{\textlangle grz\textrangle\_MAG\_SB\textlangle sblevel\textrangle\_ERR} & AB mag & Do not use; see Appendix~\ref{appendix:issues}. \\
\enddata
\tablecomments{Ellipse-fitting results and surface-brightness profiles for a
single galaxy in the \shortatlas; specifically, this table documents the data
model for the \texttt{GROUP\_NAME}-largegalaxy-\texttt{SGA\_ID}-ellipse.fits
file listed in Table~\ref{table:images}. In this table \texttt{\textlangle
grz\textrangle} denotes the $g$-, $r$-, or $z$-band filter
and \texttt{\textlangle sblevel\textrangle} represents the 22, 22.5, 23, 23.5,
24, 24.5, 25, 25.5, and 26~mag~arcsec$^{-2}$ isophote.}
\tablenotetext{a}{See the \texttt{photutils.isophote.Ellipse.fit\_image} method documentation.}
\tablenotetext{b}{See the \texttt{photutils.isophote.Isophote} and \texttt{photutils.isophote.IsophoteList} method documentation.}
\end{deluxetable}

\bibliographystyle{aasjournal}

\end{document}